\documentclass[11pt]{article}
\usepackage{jheppubmodnew}
\pdfoutput=1
\usepackage{psfrag}
\usepackage{array}
\usepackage{amssymb}
\usepackage{amsmath}
\usepackage{amsthm}
\usepackage{mathtools}
\usepackage{graphicx}
\usepackage{comment}
\usepackage[labelsep=quad]{subcaption}
	
\usepackage{cite}
\usepackage{epsfig}
\usepackage{tikz}
\usetikzlibrary{arrows,plotmarks,positioning,calc}
\usepackage{mathtools}
\newcommand{\defeq}{\vcentcolon=}
\newcommand{\eqdef}{=\vcentcolon}
\newcommand{\alset}{{\cal A}}

\newcommand{\almass}{{\cal A}^{m}}

\newcommand{\hc}{\text{h.c.}}
\newcommand{\dsthree}{{\text{dS}_3}}

\def\op{{\cal O}}

\def\pb[#1,#2]{\{#1, #2\}}
\def\deb[#1,#2]{[#1,#2]_{\text{D.B.}}}

\def\half{{1 \over 2}}

\def\Or[#1]{{\text{O}}\left({#1}\right)}
\def\dotl[#1,#2]{\left\langle #1,\, #2 \right\rangle}
\def\dotlb[#1,#2]{\left\langle #1,\, #2 \right\rangle}
\def\dotlm[#1,#2]{\left[ #1,\, #2 \right]}
\def\dotp[#1,#2]{(\vect{#1} \cdot\vect{#2})}
\def\aff[#1,#2]{\hat{#1}(#2)}

\def\n4sym{{\cal N}=4 SYM}
\def\>{\rangle}
\def\<{\langle}
\def\weight[#1,#2,#3]{\{(#1),#2,#3\}}
\def\ads[#1]{$\text{AdS}_{#1}$}

\newcommand{\be}{\begin{equation}}
\newcommand{\ee}{\end{equation}}
\newcommand{\ba}{\begin{align}}
\newcommand{\ea}{\end{align}}
\newcommand{\bs}{\begin{split}}
\def\sess\end{split}
\newcommand{\vect}[1]{{\boldsymbol{#1}}}

\def \bea {\begin{eqnarray}}
\def \eea {\end{eqnarray}}
\def \bea* {\begin{eqnarray*}}
\def \eea* {\end{eqnarray*}}

\def \bes {\begin{equation*}}
\def \ees {\end{equation*}}

\def \func{{\cal F}}
\def \orthfunc{{\cal G}}

\def\opspat{{\cal Z}}

\def \sph{{\Omega}}
\def \scri {{\cal I}}
\def \scrip{{\cal I}^{+}}
\def \scrim{{\cal I}^{-}}
\def \scrippast{{\cal I}^{+}_{-}}

\def \gfull{{\mathfrak g}}

\def \spatinf{\hat{i}^0}

\def\alcut[#1]{{\cal A}_{#1, \epsilon}}
\def\alseg[#1,#2]{{\cal B}_{#1, #2}}

\def\supcharge[#1]{\{#1\}}

\def\projsupeig[#1]{{\cal P}_{{\ell, m}}[{#1}]}
\def\transop[#1, #2]{T_{\{#1\}, \{#2\}}}
\def\supket[#1]{|\{#1\} \rangle}
\def\supbra[#1]{\langle \{#1\} | }

\def\hilbzerosect[#1]{{\cal H}^0_{\{#1\}}}
\def\hilbsect[#1]{{\cal H}_{\{#1\}}}
\def\hilbmass[#1]{{\cal H}^m}

\newcommand{\taub}{\tau}
\newcommand{\vl}{\vec{\ell}}
\def\smrf{{\mathcal{K}}}
\newcommand{\dmuxi}{d\mu_{\xi}}
\def\tB{\widetilde{B}}
\def\tN{\widetilde{N}}
\def\dds{d \mu_{X}}
\def\intpot{U}
\def\phipert[#1]{\delta \phi^{(#1)}}
\def\opspatpert[#1]{\delta \opspat^{(#1)}}
\def\phitail{\widetilde{\phi}}
\def\phitailsub[#1]{\widetilde{\phi}^{(#1)}}
\def\phinupert[#1]{\delta \phi_{\nu}^{(#1)}}

\title{Squinting at massive fields from infinity}
\author[a]{Alok Laddha,}
\author[b,c]{Siddharth G. Prabhu,}
\author[b]{Suvrat Raju}
\author[c,d]{and Pushkal Shrivastava}
\affiliation[a]{Chennai Mathematical Institute, Siruseri, Chennai, India.}
\affiliation[b]{International Centre for Theoretical Sciences, Tata Institute of Fundamental Research, Shivakote, Bengaluru 560089, India.}
\affiliation[c]{Department of Theoretical Physics, Tata Institute of Fundamental Research, Homi Bhabha Road, Mumbai 400005, India.}
\affiliation[d]{Department of Physics, Harvard University, 17 Oxford Street, Cambridge, MA 02138, USA.}
\emailAdd{aladdha@cmi.ac.in}
\emailAdd{siddharth@theory.tifr.res.in}
\emailAdd{suvrat@icts.res.in}
\emailAdd{pushkal@theory.tifr.res.in}
\date{}

\abstract{We study a novel asymptotic limit of massive scalar fields in nongravitational quantum field theories in four-dimensional flat space. We foliate the spacetime into a set of dS$_3$  slices that are spacelike to, and at a constant proper distance from, an arbitrarily chosen origin, and study the boundary dS$_3$ obtained in the infinite-distance limit.   Massive bulk fields have an exponentially small tail in this limit, and by stripping off this tail we obtain observables that are intrinsic to the boundary dS$_3$. A single massive field in the bulk can be decomposed into an infinite set of dS$_3$ fields, and the Minkowski vacuum corresponds to the Euclidean vacuum for these fields.  Our procedure for extrapolating bulk observables induces potential singularities in boundary correlators but we show how they can be cured in the free theory by smearing the boundary operators.  We show that by integrating boundary operators with suitable smearing functions it is possible to reconstruct all local bulk operators in the free theory.  We argue, using perturbation theory, that our extrapolation procedure continues to be well defined in the presence of interactions.  We  demonstrate a relationship between the width of the boundary smearing function and the localization of the bulk field.  We study other interesting properties of the boundary algebra including  the action of global translations  and the manner in which local bulk interactions are encoded on the boundary.}

\begin{document}
\maketitle

\section{Introduction \label{s_intro}}

\paragraph{\bf Motivation.}
In this paper, we study a novel limit of  bulk massive field operators in four-dimensional flat space on an asymptotic de Sitter  slice ($\dsthree$) that comprises points at infinite proper distance from a given origin. This $\dsthree$, which we denote by $\spatinf$, is related to a geometric construction called the  ``blowup'' of spatial infinity.  We show that the bulk field induces an algebra of observables on $\spatinf$, and we explore several interesting properties of this algebra.

This study originated in an attempt to better understand the principle of holography of information \cite{Laddha:2020kvp,Raju:2020smc}, which is the idea that, in any theory of quantum gravity, information that is available in the bulk of a Cauchy slice is also available near its boundary. In \cite{Laddha:2020kvp}, we showed that a state in the massless Hilbert space can be uniquely identified using observables near the past boundary of $\scrip$, or the future boundary of $\scrim$, even though, in a nongravitational theory, this would require observations on all of $\scrip$ or all of $\scrim$.

However, since observables at null infinity are insufficient to completely specify the state of massive particles, how should one generalize the result of \cite{Laddha:2020kvp} to incorporate massive particles? There is a perfectly well-defined conventional asymptotic formalism for massive fields but this involves observables at  past and future timelike infinity $i^{\pm}$ \cite{Campiglia:2015kxa}. These are the observables that enter the LSZ reduction formula for the S matrix. However, this description does not appear to be useful for understanding holography since $i^{\pm}$ are not boundaries of a Cauchy slice.  This is what led us to explore a novel asymptotic limit for massive fields in this paper.

Although the motivation for this paper comes from holography, the analysis in this paper is confined purely to {\em nongravitational} quantum-field theories in Minkowski space.  Therefore, we do not discuss flat-space holography in this paper, since holography is a property exclusively of gravitational theories.  In a forthcoming paper \cite{Laddha:2022masshol}, we will apply the formalism developed in this paper to gravitational theories. There, we will argue that, subject to the validity of physical assumptions, the principle of holography of information does generalize to theories with massive particles as follows: in nongravitational theories, such as those studied in this paper, the reconstruction of bulk observables requires data on all of $\spatinf$; but in gravitational theories, observables concentrated in a small region on $\spatinf$, when combined with gravitational measurements, can be used to completely reconstruct the bulk.

\paragraph{\bf Summary of Results.}
The metric of four-dimensional Minkowski space outside the lightcone of a given origin can be written in the form
\be
ds^2 =  d \rho^2 + \rho^2(-d \tau^2 + \cosh^2 \tau d \Omega^2),
\ee
where $\rho$ is the proper distance from the origin and each slice, at a constant value of $\rho$, is a copy of $\dsthree$.  We denote the asymptotic $\dsthree$ slice, at infinite $\rho$, by $\spatinf$ and after a suitable rescaling its metric is simply
\be
ds^2 = -d \tau^2 + \cosh^2 \tau \, d \Omega^2.
\ee
Equivalently, one may cover $\spatinf$ using coordinates $X^{\mu}$ satisfying $X^{\mu} X_{\mu} = 1$, which are related to ($\tau, \Omega$) through $X^{\mu} = (\sinh \tau, \cosh \tau \,\hat{\Omega})$. As we explain in section \ref{subsecgeometric}, the construction of $\spatinf$ is closely related to a geometric construction called ``blowing up spatial infinity''.

We argue that bulk massive field operators can be extrapolated to $\spatinf$ through the limit
\be
\opspat(\tau, \Omega) = \lim_{\rho \rightarrow \infty} D(\rho) \phi(\rho, \tau, \Omega),
\ee
with
\be
D(\rho) = \sqrt{2 \over \pi} \rho \sqrt{m \rho} e^{m \rho}.
\ee
We first derive this extrapolate dictionary by studying  the free theory but we argue that this limit also makes sense in the interacting theory when $m$ is taken to be the renormalized mass. 

In the free-field case, the boundary operator $\opspat(\tau, \Omega)$ can be decomposed into an infinite set of de Sitter component fields with mass parameter $\nu$
\be
\opspat(\tau, \Omega) = \int_0^{\infty} d \nu \phi_{\nu} (\tau, \Omega).
\ee
Moreover, from the point of view of the de Sitter component fields, the Minkowski vacuum corresponds precisely to, what is called, the Euclidean vacuum in the de Sitter literature.  In the free-field case, the boundary operator can be expanded in a basis of standard Euclidean modes, 
\be
\opspat(\tau, \Omega) = \int_0^{\infty} d\nu \,  {e^{\pi \nu \over 2} \over \sqrt{2} \, \pi} \sum_{\vec{\ell}\,}  \tB_{\nu,\vec{\ell}\,} \func_{\nu, \ell}(\tau) Y_{\vec{\ell}\,}(\Omega), 
\ee
where the precise form of $\func_{\nu, \ell}$ is specified in section \ref{subsecsphbasis}, $\tB_{\nu, \vl}$ are canonically normalized oscillators and $Y_{\vl}$ are the standard spherical harmonics.

The operators, $\opspat$, so obtained, are subtle due to the growing factor of $e^{\pi \nu \over 2}$ above that assigns exponentially increasing weight to modes at large $\nu$.   This can cause correlators of $\opspat$ to suffer from potential divergences, which can be thought of as infrared divergences since they arise only in the $\rho \rightarrow \infty$ limit.   In the free theory, we show that these divergences can be regulated by smearing the operators with an appropriate set of smearing functions
\be
\opspat(g) = \int \cosh^2 \tau d \tau d \Omega \, g(\tau, \Omega) \opspat(\tau, \Omega) . 
\ee
In the free theory, we identify the precise properties that the function $g$ must have, in order to obtain observables with finite fluctuations. This property is simply that when $g(\tau, \Omega)$ is decomposed into a set of modes dual to the modes displayed above with coefficients $\gfull_{\nu, \ell}$ so that 
\be
\opspat(g) =  \sum_{\vec{\ell\,}} \int_0^{\infty} d \nu {e^{\pi \nu \over 2} \over \sqrt{2} \, \pi} \, \tB_{\nu, \vl} \gfull_{\nu, \vl} + \hc,
\ee
then we require that $\gfull_{\nu, \ell}$ dies off fast enough at large $\nu$ to cancel the large-$\nu$ growth above and also dies off at large $\ell$.   An alternate perspective is to understand correlators of $\opspat(\tau, \Omega)$  as distributions that act on a restricted class of test functions.

We argue that the extrapolate limit is well defined in the interacting theory. Our line of argument is similar to the one used in proving the LSZ formula.  In the interacting theory, a correlator with an insertion of an operator close to the boundary is computed by a Feynman diagram where this operator is connected to the rest of the diagram with a renormalized propagator. If this renormalized propagator has a spectral decomposition with a delta function at the renormalized mass, $m$,  and no support below this value, then the falloff of this correlator is also controlled by $D(\rho)^{-1}$. Therefore, the boundary limit of the correlator is obtained by stripping off this factor. 

In correlators of the interacting theory, we find the same exponential factor that grows with the de Sitter mass parameter of the component fields. We expect that this can be handled by smearing, as in the free theory, although we do not provide a proof of this claim.

In the free theory, we show that by smearing the boundary operator with carefully chosen smearing functions, one can obtain all local bulk operators. This procedure is reminiscent of the HKLL reconstruction \cite{Hamilton:2005ju, Hamilton:2006az, Hamilton:2006fh, Hamilton:2007wj,Kabat:2011rz} in AdS/CFT \cite{Maldacena:1997re,Witten:1998qj,Gubser:1998bc}. 
For the spherically symmetric mode, we are able to obtain a nice closed form expression for the ``transfer function.'' We write
\be
	\phi(\rho_0,\tau_0, \Omega_0) = \int \cosh^2 \tau d \tau d \Omega \, \smrf(\rho_0,\tau_0, \Omega_0| \tau, \Omega) \, \opspat(\tau, \Omega),
\ee
where $\smrf$ is represented by an infinite sum given in section \ref{brobo}. Focusing on the s-wave component of the bulk and boundary operators yields a beautifully simple formula
\be
\int  \phi(\rho_0, \tau_0, \Omega_0) d \Omega_0  = \int \cosh^2 \tau d \tau Z(\tau, \Omega) \left[{1 \over 2}\frac{1}{\rho_0 \cosh{\tau_0} \cosh \tau }e^{-m\rho_0 \cosh(\tau_0-\taub)} \right] \, d \Omega . 
\ee

This boundary to bulk mapping has some robust features that hold both in the free and the interacting theory. If one considers smearing functions $g$ that die off faster than $e^{-\delta^{-1} \cosh \tau}$ for large $\tau$, then $Z(g)$ corresponds to bulk operators in the exterior of a ball of radius ${1 \over 2 m \delta}$. 
\be
[Z(g), \phi(\rho_0, \tau, \Omega)] = 0, \qquad~\text{if}~\rho_0 < {1 \over 2 m \delta}~\text{and}~e^{\delta^{-1} \cosh \tau} g(\tau, \Omega) \underset{\tau \rightarrow \pm \infty}{\longrightarrow} 0.
\ee

We also initiate a study of the transformation of operators $Z(\tau, \Omega)$ under the asymptotic symmetry group. Under global translations, by a parameter $y^{\mu}$, we show that the smearing function transforms as
\be
g(X) \rightarrow e^{-m y \cdot X} g(X), 
\ee
provided both sides of the equation above are admissible in the class of allowed smearing functions. 

Finally, we investigate perturbative corrections to these boundary operators resulting from self-interactions. We show that bulk interactions generally couple different de Sitter component fields. As an example, we show that for  a $\lambda \phi^3$ bulk interaction, the interaction Lagrangian takes the form
\be
L_{\text{int}}  =\, {\lambda} \cosh^2 \tau \int\,  \sqrt{-g_{\dsthree}} d \Omega \, \int \, f(\nu_{1}, \nu_{2}, \nu_{3})\, \phi_{\nu_{1}}(X)\, \phi_{\nu_{2}}(X)\, \phi_{\nu_{3}}(X) d \nu_1 d \nu_2 d \nu_3, 
\ee
where $f(\nu_1, \nu_2, \nu_3)$ is a specific linear combination of Appell hypergeometric functions given in section \ref{subsecinteractlag}.

\paragraph{\bf Why study observables at infinity?}

The reader might wonder why we insist on examining observables at infinity, given the subtleties of the extrapolate limit alluded to above. This is motivated by our eventual objective of applying this formalism to theories of quantum gravity. Although one integrates over bulk metrics in quantum gravity, the asymptotic part of the spacetime is still held fixed. Therefore, boundary observables remain well defined in the quantum theory and allow for a clean formulation of the physics. One can then compare properties of this boundary algebra between the gravitational and nongravitational case, as we will do for the algebra at $\spatinf$ in \cite{Laddha:2022masshol}.

However, even from a purely quantum-field theoretic perspective,  asymptotic algebras have nice properties. For example, the S-matrix is independent of field redefinitions, and the LSZ reduction formula can be applied to any field whose two-point function has a spectral representation with a delta function at the right renormalized mass \cite{Coleman:2018mew}. We show that this is also true for the limit at $\spatinf$.

\paragraph{\bf Organization of the paper.}
In section \ref{s_review}, we review the standard quantization of massive fields at $i^{\pm}$. In section \ref{s_dsfree}, we study free quantum fields in a de Sitter slicing of Minkowski space. In section \ref{subsecextrap}, we study the extrapolate limit for these observables as one approaches $\spatinf$. We describe the subtleties in this limit and explain how they can be handled in the free-field theory. We also derive results for the boundary-bulk transfer function and the transformation of fields under translations. In section \ref{s_interactions}, we argue that the extrapolate dictionary is well defined in the interacting theory using the language of Feynman diagrams, and explain which of our results remain valid in the interacting theory.  In  section \ref{secasympanal}, we describe a more conventional approach to these fields by studying the asymptotic form of the classical equations of motion. We also show how the bulk Lagrangian can be rewritten in terms of boundary operators. We end with conclusions and an outlook. The appendices contain technical details about various results that are used in the main body of the paper.

This paper builds on extensive previous work on holography in flat space. We would particularly like to refer the reader to \cite{deBoer:2003vf} --- where dS and AdS slicings of Minkowski space were explored --- and  \cite{Marolf:2006bk}, where the dynamics at the blowup of spatial infinity was explored.

\section{Review: conventional extrapolate dictionary for massive particles at $i^{\pm}$ \label{s_review}}
In this section, we briefly review the conventional approach to massive particles in an asymptotically flat spacetime.  The reader who is familiar with this material may skip directly to the next section, and we refer the reader to \cite{Campiglia:2015kxa} for a more detailed discussion. In appendix \ref{appnulldiff}, we  review the asymptotic quantization of massless particles at null infinity and explain the difficulty in representing massive particles at null infinity.

Consider an asymptotically flat metric in the vicinity of $i^{\pm}$  and introduce a coordinate $T = \sqrt{t^2 - r^2}$. The slices at constant $T$ take on the form of a Euclidean AdS$_3$, which we denote by ${\mathcal{H}_3}$, and the metric takes the form
\be\label{eq:hypslicing}
ds^2 \underset{T \rightarrow \pm \infty}{\longrightarrow} -d T^2 + T^2 d \mathcal{H}_3^2, 
\ee
where  $d \mathcal{H}_3^2$ is the line element on ${\mathcal{H}_3}$.

This coordinate system is useful for the following reason. In the usual Penrose compactification, future timelike infinity is represented by a point, $i^{+}$. But taking the limit $T \rightarrow \infty$ in the coordinate system above yields a copy of ${\mathcal H}_3$ that can be thought of as a blowup of $i^{+}$. Similarly the limit $T \rightarrow -\infty$ yields a blowup of $i^{-}$.

For $T \rightarrow \infty$,  it is expected that the fields become free, and so the behaviour of the massive Heisenberg operator is expected to be the same as the free-field behaviour,
\be
\label{phinearip}
\phi(T,\vec{p}) \underset{T \rightarrow \infty}{\longrightarrow} \frac{\sqrt{m}}{2(2 \pi T)^{3/2}}\, a_{\vec{p}}\, e^{-i m T} + \hc +O(T^{-5/2}),
\ee
where the momentum, $\vec{p}$, provides a natural coordinate on the ${\mathcal{H}_3}$ at $T \rightarrow  \infty$.
In \cite{Campiglia:2015kxa}, it is shown that the operators $a_{\vec{p}}$ obey the commutation relations
 \be
\label{commutrel}
[a_{\vec{p}}, a^{\dagger}_{\vec{p'}}]= 2\, (2 \pi)^3 \, E_{p} \, \delta^{(3)}(\vec{p}-\vec{p'})
 .\ee
Note that the commutators \eqref{commutrel} are exact, even in the full theory, since they pertain to operators defined at asymptotic infinity. 

The map from bulk operators to the operators $a_{\vec{p}}, a_{\vec{p}}^{\dagger}$ that are intrinsic to the blowup of $i^{+}$ is the conventional extrapolate dictionary for massive particles.

A similar extrapolate dictionary holds for the Heisenberg operator near $i^{-}$ which can be blown up to yield a {\em separate} hyperboloid. With $\vec{p}$ now denoting coordinates on the hyperboloid at the blowup of $i^{-}$, we have 
\be
\label{phinearimin}
\phi(T,\vec{p}) \underset{T \rightarrow -\infty}{\longrightarrow} \frac{\sqrt{m}}{2(2 \pi |T|)^{3/2}}\, \tilde{a}_{\vec{p}}\, e^{-i m T} + \hc +O(T^{-5/2}),
\ee
and the operators that appear here also obey $ [\tilde{a}_{\vec{p}},\tilde{a}^{\dagger}_{\vec{p'}}]= 2\, (2 \pi)^3 \, E_{p} \, \delta^{(3)}(\vec{p}-\vec{p'})]$. 

The operators $\tilde{a}_{\vec{p}}$
that appear in \eqref{phinearimin} are not the same as the operators $a_{\vec{p}}$ that appear in \eqref{phinearip}. The action of the $\tilde{a}_{\vec{p}}$ algebra on the vacuum generates the ``in''-states. The action of the  operators $a_{\vec{p}}$ generates the ``out''-states. Although both algebras are free, and both the ``in'' and ``out'' Hilbert space are Fock spaces, the mapping between them is nontrivial and is given by the S matrix.

The algebra of massive operators is generated by all possible polynomials in the $a_{\vec{p}}$ operators and their conjugates
\be
\almass = \text{span}\{ a^{\dagger}_{\vec{p}_{1}} \ldots a^{\dagger}_{\vec{p_n}} a_{\vec{p}_{n+1}}  \ldots a_{\vec{p_m}}\}. 
\ee
A similar algebra may be obtained from polynomials in the $\tilde{a}_{\vec{p}}$ operators.  In a theory where there are no other fields, the two algebras are isomorphic by unitarity but, in general, they are distinct.

The action of this algebra on the vacuum gives the Hilbert space of massive particles.
\be
\label{hmassiveconstruct}
{\cal H}^{m} = \almass | 0 \rangle.
\ee
The full Hilbert space of the theory, that encompasses both massive and massless excitations, is simply the direct product
\be
\label{hfulldef}
{\cal H} = {\cal H}^{m} \otimes {\cal H}^{0},
\ee
where ${\cal H}^{0}$ is the Hilbert space of massless particles defined at $\scrip$. (See appendix \ref{edmfao} for more details.) 

Note that the Hilbert space has a natural product basis when it is written in terms of massless particles on $\scrip$ and massive particles on $i^{+}$. Correspondingly, another natural product basis is provided in terms of states of massless particles on $\scrim$ and massive particles on $i^{-}$.  On the other hand, operators on $\scrip$ do not commute with operators on $i^{-}$ or on ${\scrim}$ and vice versa. 

The conventional extrapolate dictionary for massive operators is perfectly satisfactory for the purpose of defining an S-matrix. However, from the point of view of holography, it appears to be more useful to have a description of massive fields near the boundary of a Cauchy slice rather than in the asymptotic future or the asymptotic past. For this reason, we now turn to a novel asymptotic description of massive particles.

\section{Quantum field theory in a de Sitter slicing  \label{s_dsfree}}
In this section, we develop some tools to study massive fields on a de Sitter slicing of Minkowski space. This section is a preparatory section for eventually understanding the algebra of observables on $\spatinf$.  We will study a free field in empty Minkowski space that obeys the equation of motion 
\be
\label{freeeom}
(\Box - m^2) \, \phi = 0. 
\ee
We will study this field  in a coordinate system that utilizes a de Sitter slicing  of Minkowski space outside the light cone.  A similar analysis was performed in \cite{deBoer:2003vf, Marolf:2006bk}. 

\begin{figure}[!h]
	\begin{center}
		\includegraphics[height=0.4\textheight]{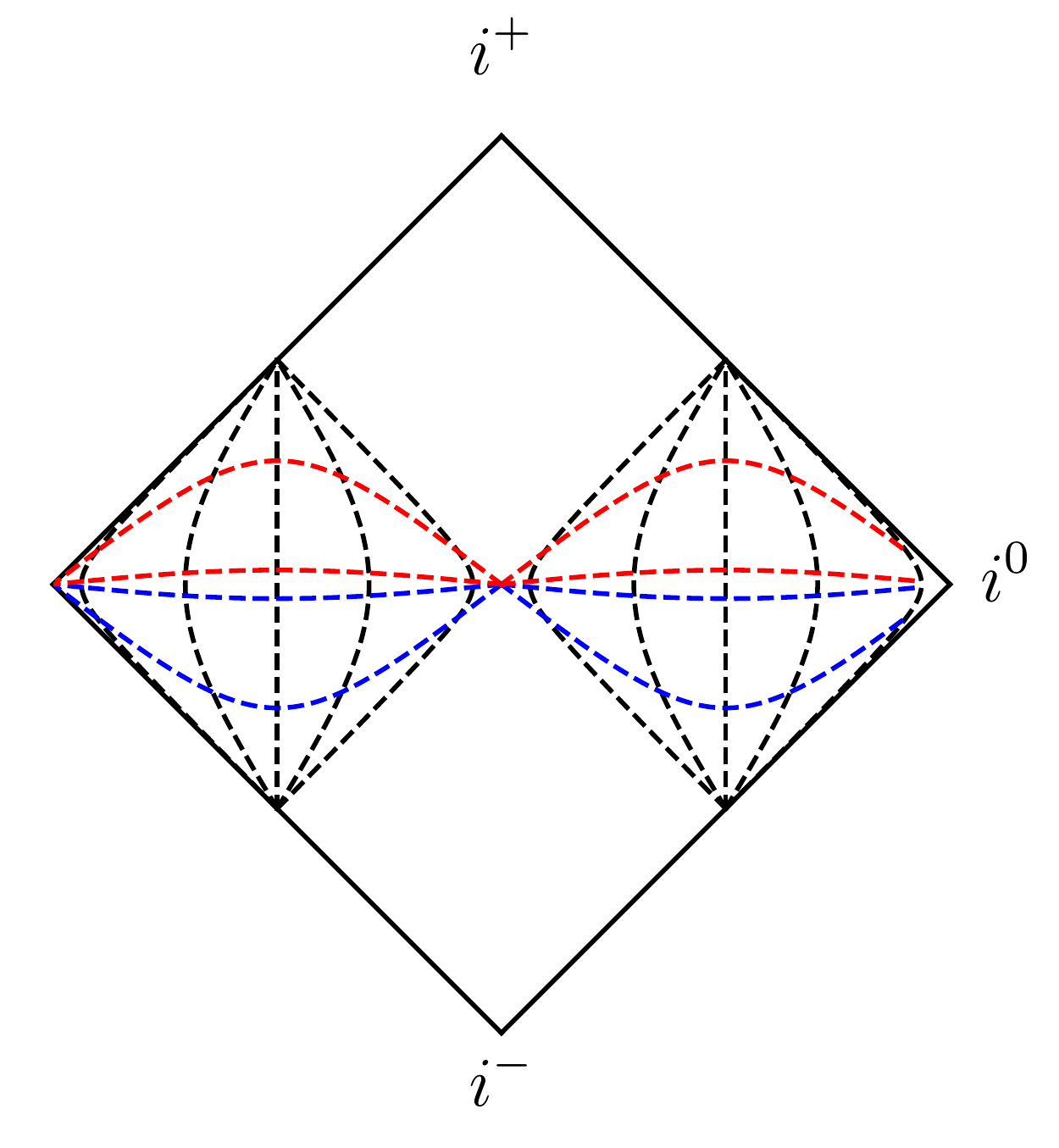}
		\caption{\em A figure depicting $(\tau, \rho)$ coordinates in Minkowski space. Lines of constant $\rho$ are black dashed lines. Lines of constant positive $\tau$ are red dashed lines, while lines of constant negative $\tau$ are shown as blue dashed lines. \label{figrhotau}}
	\end{center}
\end{figure}

Starting with the metric of Minkowski space written as
\be
ds^2 =  \eta_{\mu \nu} d x^{\mu} d x^{\nu} = -d t^2 + d r^2 + r^2 d \Omega^2, 
\ee
we introduce coordinates  $\rho, \tau$ outside the light cone so that
\be
\label{hypcoords}
t = \rho \sinh \tau; \qquad r = \rho \cosh \tau.
\ee
In these coordinates the metric of Minkowski space becomes
\be
\label{rhotaucoords}
ds^2 =  d \rho^2 + \rho^2(-d \tau^2 + \cosh^2 \tau d \Omega^2). 
\ee
The $(\rho, \tau)$ coordinates are displayed in Figure \ref{figrhotau}.

We recognize that at each value of $\rho$, we have a copy of 3-dimensional de Sitter space. A convenient set of coordinates on $\dsthree$ is given by unit  vectors satisfying
\be
\eta_{\mu \nu} X^{\mu} X^{\nu} = 1.
\ee
In the ambient Minkowski space, these vectors can be identified with $X^{\mu} = x^{\mu}/\rho$.  We will usually omit the superscript on $X$ when we use the coordinate to specify the location of field operators like $\phi(\rho, X)$. Note that $X$ is a four-vector and in the  global coordinates above, it is given by
\be
\label{xtotauomega}
X = (\sinh \tau, \cosh \tau \ \Omega_x, \cosh \tau \ \Omega_y, \cosh \tau \ \Omega_z), 
\ee
where $(\Omega_x, \Omega_y, \Omega_z)$ is a unit 3-vector that specifies a position on the $S^2$.

In these coordinates, the solution of \eqref{freeeom} takes the form
\be
\label{dsdecomp}
\phi(\rho, X)\, =  \int_0^{\infty} d \nu  {1 \over \rho} K_{i \nu}(m \rho) \phi_{\nu}(X).
\ee
We have put a normalizability condition on $\phi$ in \eqref{dsdecomp} and discarded solutions that grow exponentially at large $\rho$. The modified Bessel function of imaginary order, $K_{i \nu}$, is also called the Macdonald function. In \eqref{dsdecomp},  $X$ is a de Sitter coordinate as above and  $\phi_{\nu}(X)$ is a de Sitter free field that satisfies
\be
\label{dseqn}
(\Box_{\dsthree} - (1 + \nu^2)) \phi_{\nu}(X) = 0.
\ee
This equation of motion coincides with that of a massive free field on $\textrm{dS}_{3}$.

We now proceed to quantize the one parameter family of massive fields $\phi_{\nu}(X)$ on $\textrm{dS}_{3}$ such that the spectral integral $\int\, d\nu\, \frac{K_{i\nu}(m\rho)}{\rho}\, \phi_{\nu}(X)$ with a  measure factor $\frac{K_{i\nu}(m\rho)}{\rho}$ is the bulk scalar field operator. As we will see,  the choice of $\textrm{dS}_{3}$ vacuum that corresponds to the Minkowski vacuum is precisely the Euclidean vacuum. We will perform this quantization in two bases of solutions to \eqref{dseqn} --- a ``conformal basis'' and a ``spherical basis''.

\subsection{Conformal basis} \label{confbasis}
The first basis that we describe is often called the principal basis \cite{Bros:1995js} in the dS literature. However, we adopt the terminology ``conformal basis'' to emphasize the similarity with a similar basis for a Euclidean-AdS slicing of Minkowski space introduced in \cite{Pasterski:2016qvg, Pasterski:2017kqt}. (See also \cite{Melton:2021kkz}.) We caution the reader that the basis we describe is not identical to the basis of \cite{Pasterski:2016qvg, Pasterski:2017kqt}.  The authors of \cite{Pasterski:2016qvg, Pasterski:2017kqt} were interested in expanding the field in the forward and past light cone whereas we are considering the region outside the light cone. The two bases are related by a Bogoliubov transformation that, in an interacting theory, depends on the details of the interactions.

\paragraph{Mode functions and analyticity properties.}

Let $\xi$ be a future-directed null vector. Consider the mode function $\psi_{\nu,\xi}(X) = \frac{1}{(-X\cdot \xi)^{1 + i\nu}}$ with $\nu > 0$,  which we initially define in the regime where $X \cdot \xi < 0$.   Note that scaling $\xi$ does not produce a new solution.  Then it is easy to check that $\psi_{\nu, \xi}$ and its conjugate satisfies the de Sitter wave equation \eqref{dseqn}.  We will now extend the validity of this mode function to all $X$, and find that these mode functions and their conjugates, when parameterized by all possible choices of $\nu$ and $\xi$, provide a complete basis.

To extend the domain of the mode function to all $X$, we need to specify its analyticity properties. We demand that the mode function $K_{i\nu}(m\rho) \psi_{\nu,\xi}/\rho$ be writable as a combination of only positive energy Minkowski plane wave modes, so that the operators associated with these modes in the decomposition of the scalar field, annihilate the Minkowski vacuum. The positive energy Minkowski modes $e^{-i E_{\vec{p}\,} t + i \vec{p}\cdot \vec{x}}$ are analytic under $x^{\mu} \rightarrow x^{\mu}+i y^{\mu}$ if $y$ is a past-directed time-like vector. If the conformal modes share this analyticity property, then a simple contour-deformation argument shows that the Klein-Gordon inner product of  $K_{i\nu}(m\rho) \psi_{\nu,\xi}/\rho$ with negative energy plane wave modes vanishes.

We therefore require that the mode function $\psi_{\nu,\xi}(Z)$ should be analytic in the following complex domain, called ``the past tube'' in the de Sitter literature.
\be
\label{pasttubeext}
 \{Z = X+i Y \ | \  Z^2=1 \implies X^2-Y^2=1, \, X \cdot Y=0 ; \ Y^2<0 \text{ and } Y^0 < 0\}. 
\ee
We can achieve this by introducing the following $i\epsilon$-prescription in the definition of the mode function: 
\be\label{modefn}
\psi_{\nu,\xi}(X) \defeq {1 \over (-X\cdot \xi - i \epsilon)^{1 + i\nu}} = {\theta(-X\cdot\xi) - \theta(X\cdot\xi) \, e^{-\pi \nu} \over |X\cdot \xi|^{1 + i\nu}} .
\ee
The $i\epsilon$ prescription fixes the coefficient of $\theta(X\cdot \xi)$. It also ensures that when $X$ is extended to $Z$ in the past tube \eqref{pasttubeext},  the imaginary part of $-Z\cdot \xi$ is negative ensuring that the mode function is analytic in the past tube. 

Similarly, we can define the negative energy modes, which are simply the conjugate of the positive energy modes.
\be\label{modefnnegative}
\bar{\psi}_{\nu,\xi}(X)  = {1 \over (-X\cdot \xi + i \epsilon)^{1 - i\nu}} = {\theta(-X.\xi) - \theta(X.\xi) \, e^{-\pi \nu} \over |X\cdot \xi|^{1 - i\nu}}.
\ee
The negative energy modes are analytic in the future tube which corresponds to the extension
\be
\label{futuretubeext}
 \{Z = X+i Y \ | \  Z^2=1 \implies X^2-Y^2=1, \, X \cdot Y=0 ; \ Y^2<0 \text{ and } Y^0 > 0\}. 
\ee

\paragraph{Mode expansion of the massive scalar field.} To quantize the field, we simply expand it as 
\be\label{modeexpconfp}
	\phi(\rho,X) =  \int_0^{\infty} d\nu \, \dmuxi \, N_{\nu} \, B_{\nu,\xi} \,  {K_{i\nu}(m \rho) \over \rho } \, \psi_{\nu,\xi}(X) + \hc , 
\ee
where, $\dmuxi$ is a volume form on the space of inequivalent future-directed null vectors  \cite{Bros:1995js} (see appendix \ref{app_moredetails} for an explicit expression) ; and in four dimensional spacetime, the normalization factor
\be
\label{planenorm}
N_{\nu} = {\nu e^{\pi \nu \over 2} \over 2^{3 \over 2} \pi^2}, 
\ee
ensures that when $B_{\nu, \xi}, B_{\nu, \xi}^{\dagger}$ are promoted to oscillators, these oscillators are canonically normalized 
\be\label{bbdaggercomm}
[B_{\nu, \xi}, B^{\dagger}_{\nu', \xi'}] = \delta(\nu - \nu') {\delta^2(\xi, \xi')}.
\ee
  The normalization is derived in detail in Appendix \ref{app_confbasis}.

\paragraph{Hilbert space.}
In free-field theory, a Fock space is obtained by starting with the vacuum annihilated by all the $B_{\nu, \xi}$
\be
\label{bnuannihilateszero}
B_{\nu, \xi} | 0 \rangle = 0.
\ee
If one focuses on a de Sitter component field with a given $\nu$, then \eqref{bnuannihilateszero} also defines what is called the Euclidean vacuum (alternately called the Bunch-Davies vacuum) in de Sitter space. 
The space of massive excitations is then obtained via
\be
\label{hilbconformal}
{\cal H}^m = \text{span}\{B_{\nu_1, \xi_1}^{\dagger} \ldots B_{\nu_n, \xi_n}^{\dagger} | 0 \rangle \},
\ee
which, in free-field theory, manifestly coincides with \eqref{hmassiveconstruct} since there is a Bogoliubov transformation that ensures that the oscillators at $i^{+}$ (i.e $a_{\vec{p}},\, a_{\vec{p}}^{\dagger}$ displayed in \eqref{phinearip} ) are unitarily related to the $B_{\nu, \xi}, B_{\nu, \xi}^{\dagger}$ operators.

\subsection{Spherical basis \label{subsecsphbasis}}
\paragraph{Mode expansion.} The topology  of $\dsthree$  is $S^{2}\, \times\, R$. This motivates the use of, what we call, the ``spherical basis.'' The spherical basis utilizes the  spherical harmonics of the $S^2$ and sometimes turns out to be more convenient from a computational perspective.

In the spherical  basis, we expand the field in Minkowski space as
\be
\label{sphbasis}
	\phi(\rho,X) =  \int_0^{\infty} d\nu \,  {K_{i\nu}(m \rho) \over \rho } \, {\tN_{\nu} } \sum_{\vec{\ell}\,}  \tB_{\nu,\vec{\ell}\,} \func_{\nu, \ell}(\tau) Y_{\vec{\ell}\,}(\Omega) + \hc, 
\ee
where $Y_{\vec{\ell}\,}(\Omega)$ are the standard spherical harmonics on the sphere, $\vec{\ell} = (\ell,m)$ denotes the collection of all angular momentum quantum numbers, $F_{\nu,\ell}(\tau)$ is the solution to
\be\label{eq:taueq}
\left(\partial^2_\tau +2 \tanh\tau \, \partial_\tau +\ell(\ell+1)\text{sech}^2 \tau \, +1+\nu^2\right) F_{\nu,\ell}(\tau)=0,
\ee
and 
\be
\label{normeuclidean}
	\tN_{\nu} = {e^{\pi \nu \over 2} \over \sqrt{2} \, \pi}.
\ee

The solutions that serve as good positive and negative frequency modes in Minkowski space correspond to the so-called ``Euclidean modes'' in de Sitter space and their conjugates \cite{Mottola:1984ar, Allen:1985ux}. This name comes from their property that when $\dsthree$ is continued to Euclidean $S^3$, the Euclidean modes are regular on the south pole of the $S^3$ whereas their conjugates are regular on the north pole.  
More precisely, the positive frequency modes are fixed by the condition \cite{Mottola:1984ar,Allen:1985ux}
\be
\label{euclideanmodecond}
\lim_{\tau \rightarrow -{i \pi \over 2}} (\tau + i {\pi\over 2})  \func_{\nu, \ell}(\tau) = 0. 
\ee

The negative frequency modes, which are complex conjugates of the modes above, are regular in the limit $\tau \rightarrow {i \pi \over 2}$. 
As explained in more detail in appendix  \ref{appsphbasis}, we get
\be
\label{sphericalbasismodefunc}
\begin{split}
	\func_{\nu,\ell}(\tau) = {e^{\ell \tau} \over 2^{\ell} (\cosh \tau)^{\ell+1}} \left[e^{-i \nu \tau}  \, _2F_1\left(i \nu - \ell, -\ell, 1 + i \nu, -e^{-2 \tau} \right) \right.  \hspace{20 pt} \\
	 \left. + c_{\nu, \ell} e^{i \nu \tau} \, _2F_1\left(-i \nu - \ell, -\ell, 1 - i \nu, -e^{-2 \tau} \right) \right].
\end{split}
\ee
The hypergeometric functions are finite polynomials since $\ell$ is an integer. The constant 
\be
c_{\nu, \ell} =-e^{-\pi  \nu } \frac{ \Gamma (i \nu +1) \Gamma (\ell-i \nu +1)}{\Gamma (1-i \nu ) \Gamma (\ell+i \nu +1)} , 
\ee
ensures that the condition \eqref{euclideanmodecond} is met.

The normalization factor \eqref{normeuclidean} ensures that the $\tB_{\nu, \vec{\ell}}$ satisfy
\be \label{btbtdagger}
[\tB_{\nu, \vec{\ell}}, \tB^{\dagger}_{\nu', \vec{\ell\,}'}] = \delta(\nu - \nu') \, \delta_{\vec{\ell} ,\vec{\ell\,}'}. 
\ee
Further details of how the normalization is computed are provided in appendix \ref{appsphbasis}. 

There are two ways to understand why the Euclidean modes are the correct positive frequency modes. The first is that it is possible to  Wick rotate Minkowski space to Euclidean space by rotating $\tau$ in \eqref{hypcoords}. Then it is natural that the positive-frequency modes are those that are regular in the infinite past.   As an independent check, in appendix \ref{conftospheranalyticity}, we directly perform a Bogoliubov transformation of the conformal modes, and show that this leads us to the Euclidean modes.

In de Sitter quantum field theory, the operators that multiply the Euclidean modes annihilate the Euclidean vacuum, which is also obtained by doing the Euclidean path-integral on the complexified dS up to $\tau = 0$.   There has been some discussion in the literature on the choice of alternate vacua  in de Sitter space. However, here the choice of vacuum is unambiguously dictated by the ambient Minkowski vacuum.

\paragraph{Orthogonality relations.}
In appendix \ref{appsphbasis} we show that the mode functions $\func_{\nu,\ell}$, defined for $\nu>0$, satisfy the following orthogonality relations,
\be
\begin{split}
\int_{-\infty}^{\infty} \func_{\nu',\ell}(\tau) \func^*_{\nu,\ell}(\tau) \, \cosh^2\tau \, d\tau 
&= 2 \pi (1+|c_{\nu,\ell}|^2) \delta(\nu-\nu'),\\
\int_{-\infty}^{\infty} \func_{\nu',\ell}(\tau) \func_{\nu,\ell}(\tau) \, \cosh^2\tau \, d\tau 
&= 4 \pi \, c_{\nu,\ell} \,\delta(\nu-\nu').
\end{split}
\ee
Note that $|c_{\nu, \ell}|^2 = e^{-2 \pi \nu}$.
For later use, it is convenient to define a function $\orthfunc_{\nu,\ell}$ which has the property that it is orthogonal to all $\func^*_{\nu'\ell}$ and gives a non-zero integral only with $\func_{\nu, \ell}$,
\be\label{orthfunc}
\orthfunc_{\nu,\ell} (\tau) = {1+|c_{\nu,\ell}|^2 \over 2 \pi ( 1 - |c_{\nu,\ell}|^2)^2 } \, \func^*_{\nu, \ell} (\tau) - {c^{*}_{\nu,\ell} \over \pi (1 - |c_{\nu,\ell}|^2)^2 } \, \func_{\nu, \ell} (\tau), 
\ee
such that, for $\nu > 0$ and $\nu' > 0$, 
\be\label{orthognalityfuncorthfunc}
\begin{split}
\int_{-\infty}^{\infty} \orthfunc_{\nu',\ell}(\tau) \func_{\nu,\ell}(\tau) \, \cosh^2 \tau \, d\tau  &= \delta(\nu -\nu'),\\
\int_{-\infty}^{\infty} \orthfunc_{\nu',\ell}(\tau) \func^*_{\nu,\ell}(\tau) \, \cosh^2 \tau \, d\tau  &= 0.
\end{split}
\ee

\subsection{Transformation between the two bases \label{sectransform}}
In appendix \ref{appsphbasis}, it is shown that the conformal basis can be mapped precisely to the spherical basis, using the following identity
\be
{1 \over (-X\cdot\xi - i \epsilon)^{1+i\nu}} = \sum_{\vl} e^{i {\mathfrak{b}(\nu, \ell)}} {\tN_{\nu} \over N_{\nu}} \, \func_{\nu,\ell}(\tau) \, Y_{\vec{\ell\,}} (\hat{\xi}) Y_{\vl}(\Omega),
\ee
where, by $\hat{\xi}$, we mean the unit-vector on the sphere that is parallel to the spatial part of $\xi$ i.e. $\xi = (1, \hat{\xi})$, and $e^{i {\mathfrak{b}(\nu, \ell)}}$ is an unimportant phase factor given in \eqref{appphasefac}.
Therefore the positive-frequency conformal modes map into the positive-frequency spherical modes and vice versa.

Comparing \eqref{modeexpconfp} and \eqref{sphbasis}, we see that the sets of modes are related simply by
\be
\tB_{\nu, \vec{\ell}} =   e^{i {\mathfrak{b}(\nu, \ell)}} \int  \dmuxi  B_{\nu, \xi} Y_{\ell}(\hat{\xi}). 
\ee

\paragraph{Hilbert space.}
From the transformation above, it is clear that the Hilbert space of massive particles can also be obtained by specifying the vacuum through
\be
\tB_{\nu, \vec{\ell}}| 0 \rangle = 0,
\ee
and then constructing
\be
\label{hilbmassive}
{\cal H}^m = \text{span}\{\tB_{\nu_1, \vec{\ell}_1}^{\dagger} \ldots \tB_{\nu_n, \vec{\ell}_n}^{\dagger} | 0 \rangle \}.
\ee

\subsection{Two-point function}
The two-point function of  de Sitter fields with a mass parameter $\nu$ in the Euclidean vacuum   is known to be \cite{Bros:1995js}
\be
\label{twopointnu}
\begin{split}
\langle\, 0| \phi_{\nu}(X_1) \phi_{\nu'}(X_2)\, | 0 \rangle\, &=  \int d \nu \dmuxi  |N_{\nu}|^2 {1 \over (-X_1 \cdot \xi + i \epsilon)^{1 - i \nu}} {1 \over (-X_2 \cdot \xi - i \epsilon)^{1 + i \nu}}  \\ &=  {-i \over 2 \pi^3} \nu \, C^{1}_{i \nu - 1}(-X_1 \cdot X_2) \delta(\nu - \nu'),
\end{split}
\ee
where $C^1_{i \nu - 1}$ is the Gegenbauer function. This function can be represented alternately using the identities (See appendix A of \cite{Marolf:2010zp})
\be
\label{dstwopt}
C^{1}_{i \nu - 1}(z) =\, i \nu \,   _2F_1(1 -i \nu, 1+ i \nu; {3 \over 2}; {1 - z \over 2}). 
\ee
When the argument is in the range $1\, \leq\, z\, <\, \infty $, we can also write $C^{1}_{i \nu - 1}(z)$ in terms of elementary functions as
\be
\label{gegenelem}
C^{1}_{i \nu-1}(\cosh q) = i {\sin (\nu q) \over \sinh q}.
\ee

The $\textrm{dS}_{3}$ two point function \eqref{dstwopt} is consistent with the two-point function for a massive field \cite{bogoliubov1959itq}. 
This follows from the identity (see formula 19a in \cite{durand1979addition})
\be
\label{twopoint}
\begin{split}
\langle 0| \phi(\rho_1, X_1) \phi(\rho_2, X_2)|0 \rangle &= {-i \over 2 \pi^3} \int_0^{\infty} \nu d \nu {K_{i \nu}(m \rho_1) K_{i \nu}(m \rho_2) \over \rho_1 \rho_2} C^{1}_{i \nu - 1}(-X_1 \cdot X_2) \\
&= {m \over 4 \pi^2} {K_1(m w) \over  w},
\end{split} 
\ee
where 
\be
w = \sqrt{\rho_1^2 + \rho_2^2 - 2 \rho_1 \rho_2 X_1 \cdot X_2}. 
\ee
The identity is valid only for $w^2 > 0$ but the  two-point function can be analytically continued to the case where the points are time-like separated. We are primarily interested in the two-point Wightman correlator, obtained after extending $X_1$ by a small amount in the past tube (equation \eqref{pasttubeext}), and extending $X_2$ by a small amount in the future tube (equation \eqref{futuretubeext}).  

However, if instead of this $i \epsilon$ prescription, one were to take $X_1 \cdot X_2 \rightarrow X_1 \cdot X_2 \pm i \epsilon$, then one would obtain the time-ordered and anti-time-ordered correlator respectively.

\section{Extrapolate dictionary and boundary observables for free fields \label{subsecextrap}}
In section \ref{s_review}, we reviewed the conventional treatment of massive particles at timelike infinity. We now explain our construction of $\spatinf$, which is simply the de Sitter slice that lives at $\rho \rightarrow \infty$ in the foliation described previously, and its relationship to the standard geometric construction of blowing up spatial infinity. We then show that a bulk massive field induces an algebra of observables on $\spatinf$. The basic idea is simply that at large $\rho$,  the  field operator $\phi(\rho, X)$ asymptotes to an operator $\sqrt{\pi \over 2} {e^{-m\rho} \over \rho \sqrt{m \rho} } \, \opspat(X)$. This can be used to define an operator $\opspat(X)$ that is intrinsic to $\spatinf$.  However, this limit is subtle and has several interesting properties that we describe in this section.

In this section, we continue to discuss free-field theory in Minkowski space. This is because we would like to introduce the extrapolate dictionary in a simple setting. In the next section, we will generalize our discussion to the interacting theory.

\subsection{Geometric preliminaries \label{subsecgeometric}}
In the conventional Penrose diagram, spatial infinity, denoted by $i^0$ is just a point. In order to resolve the structure of $i^0$, \cite{Ashtekar:1978zz} introduced a ``blowup'' of spatial infinity that we now review. Two different representations of this blowup have been used in the literature --- one is called the ``hyperboloid at spatial infinity'' introduced by Ashtekar and Romano \cite{Ashtekar:1991vb} whereas an equivalent representation \cite{Mohamed:2021yzf}, the ``cylinder at spatial infinity'',   was described by Freidrich \cite{Friedrich:1999ax,Friedrich:1999wk,Friedrich:2002ru,Friedrich:2006km}.

\begin{figure}
	\begin{center}
		\includegraphics[height=0.4\textheight]{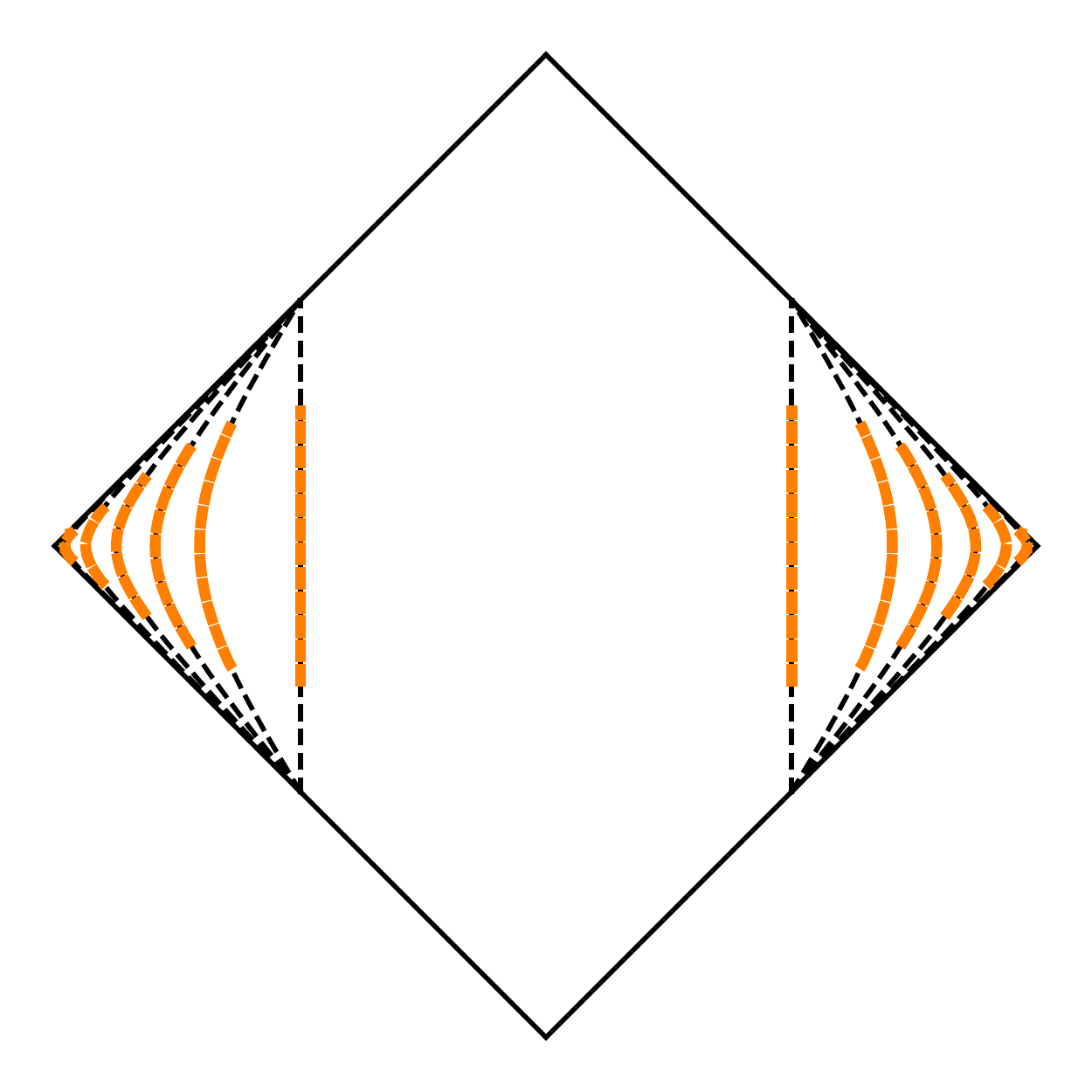}
		\caption{\em The blowup of spatial infinity. The thick orange lines show an interval that covers a fixed range of  $\tau$ (in coordinates from \eqref{rhotaucoords}) on hyperboloids of different $\rho$. As $\rho$ becomes larger, this interval moves closer  to $i^0$.\label{blowupplot}}
	\end{center}
\end{figure}

In the hyperboloid representation, the blowup of spatial infinity is directly related to the de Sitter slicing outside the light cone discussed in section \ref{s_dsfree}. It is obtained by taking the limit
\be
\rho \rightarrow \infty,
\ee
and rescaling the metric on constant $\rho$ slices by a factor of ${1 \over \rho^2}$ to obtain the three dimensional metric on the blowup of $i^0$ (see Figure \ref{blowupplot}),
\be
\label{blowupmetric}
ds_3^2 = -d \tau^2 + \cosh^2 \tau d \Omega^2.
\ee

We also use the symbol $\spatinf$ to refer to the $\dsthree$ at infinite $\rho$  with metric \eqref{blowupmetric}. However, there is a subtle difference between our usage and the perspective of \cite{Ashtekar:1978zz, Ashtekar:1991vb}.  This subtlety is depicted in Figure \ref{blowupplot}. If we focus on a fixed interval in the $(\tau, \Omega)$ coordinates and then consider a sequence of increasing values of $\rho$ this leads to a sequence of intervals that eventually become spacelike separated from any point in the bulk.  This is the limit usually considered in the geometric literature.

However, in our discussion below we will also consider operators that are smeared over {\em all} of $\dsthree$ at finite $\rho$ and then take the $\rho \rightarrow \infty$ limit. But, as Figure \ref{blowupplot} shows, at every value of $\rho$, some parts of the $\dsthree$ slice  are causally connected to points in the bulk. Consequently, the limit of such a smeared operator might not commute with operators in the bulk. So, to avoid confusion, the reader should think of $\spatinf$ simply as the asymptotic $\dsthree$ slice rather than strictly as a resolution of spatial infinity.

\subsection{Pointwise extrapolate dictionary \label{subsecptwise}}
When $\rho$ becomes arbitrarily large, we may take the limit of the Bessel functions in the expression \eqref{dsdecomp} for any $\nu$ using the formula
\be
\label{largerholimit}
K_{i \nu}(m \rho) \underset{\rho \rightarrow \infty}{\longrightarrow} \sqrt{\pi \over 2} \sqrt{1 \over m \rho} e^{-m \rho}, \qquad m \rho \gg \nu. 
\ee
This suggests the following extrapolate dictionary. We multiply the field operators  $\phi(\rho, X)$ by the factor
\be
D(\rho) = \sqrt{2 \over \pi} \rho \sqrt{m \rho} e^{m \rho},
\ee
and formally take the large $\rho$ limit to obtain operators that are intrinsic to $\spatinf$
\be
\label{extrapolatedict}
\opspat(X) = \lim_{\rho \rightarrow \infty} D(\rho) \phi(\rho, X) = \int \phi_{\nu}(X) d \nu.
\ee
In terms of the creation and annihilation operators discussed above
\be
\label{extrapolosc}
\begin{split}
	\opspat(X) &= \int d \nu \, \dmuxi \, N_{\nu}  \, {B_{\nu, \xi}  \over (-X.\xi - i \epsilon)^{1 + i \nu}} + \hc \\ 
	&= \sum_{\vec{\ell}\,} \int d \nu \, \tN_{\nu} \,  \tB_{\nu, \vec{\ell}} \, \func_{\nu, \ell}(\tau) \, Y_{\vl}(\Omega) + \hc
\end{split}
\ee

\subsection{Subtlety in the pointwise extrapolate limit \label{subsecsubtlety}}

The pointwise limit displayed in \eqref{extrapolatedict} must be treated with care.  The creation and annihilation operators that appear in \eqref{extrapolosc} are canonically normalized but the leading normalization factors scale exponentially with $\nu$ as can be seen from the factor of $e^{\pi \nu \over 2}$ in \eqref{planenorm} and \eqref{normeuclidean}. Consequently, the two-point function in the vacuum of two pointwise extrapolated operators might not be well defined. Using the two-point function \eqref{twopointnu} we find
\be
\langle 0 | \opspat(X_1) \opspat(X_2) | 0 \rangle = {-i \over 2 \pi^3} \int_0^{\infty} \nu d \nu  C^{1}_{i \nu - 1}(-X_1 \cdot X_2) \quad \text{is~undefined~for}~X_1 \cdot X_2 > -1,
\ee
because the integral over $\nu$ does not converge as can be seen from the expression \eqref{gegenelem}.  On the other hand, for $X_1 \cdot X_2 < -1$ the two-point function vanishes.

This can also be seen directly by going back to the bulk expression for the two-point function \eqref{twopoint}. We find that 
\be
\label{twoptdiverg}
\lim_{\rho \rightarrow \infty} D(\rho)^2 \langle 0 | \phi(\rho, X_1) \phi(\rho, X_2) |0 \rangle \quad \text{does~not~exist~if~} X_1 \cdot X_2 > -1,
\ee
since in the regime above the Bessel function on the right hand side of \eqref{twopoint} decays {\em slower} than $e^{-2 m \rho}$. For $X_1 \cdot X_2 < -1$, the limit vanishes.

This problem can be thought of as an infrared issue since it occurs only when we take both the points in \eqref{twoptdiverg} to infinity. If we keep either of the points at a finite value of $\rho$ then the Bessel function that appears in \eqref{dsdecomp} acts as a regulating factor for the large $\nu$ components of the field. For any finite $\rho$, in the limit  where $\nu$ becomes arbitrarily large \cite{dunster1990bessel}
\be
\label{largenulimit}
K_{i \nu}(m \rho) \underset{\nu \rightarrow \infty}{\longrightarrow} \left(\pi \over \nu \sinh \pi \nu\right)^{1 \over 2} \sin \left(-\nu  \log \left(\frac{m \rho}{2}\right)+\arg (\Gamma (i \nu +1))\right), 
\ee
which decays as $e^{- \nu \pi \over 2}$ at large $\nu$. (Note that there is no contradiction with \eqref{largerholimit} since in \eqref{largenulimit}, we assume that $\nu \gg m \rho$.)\footnote{A more detailed large-parameter analysis of \ref{dsdecomp} can be found in \cite{Temme1994}. We thank Semyon Yakubovich for pointing out this reference to us.}

However,  the exponential suppression factor at large $\nu$ supplied by the Bessel function that is present in \eqref{dsdecomp} has been removed in the operators \eqref{extrapolatedict}.  This is what leads to the subtlety in the correlators described above.  We will show how this issue can be addressed by smearing the operators to obtain observables that have finite correlators and can be multiplied to obtain an algebra of operators at $\spatinf$.

Before we discuss the smearing, we note that, in some circumstances, the pointwise operators defined above might make sense inside correlation functions. For instance, let $|\Omega \rangle$ be a normalizable state from \eqref{hilbconformal} with bounded ``$\nu$-content'' and ``$\ell$-content''. By this, we mean that it is formed by the action of creation operators with a {\em bounded} value of $\nu$ and $\ell$ on the vacuum.  (A slightly weaker condition can be placed on $|\Omega \rangle$, which will become clear below.) Then  the expectation value of $\phi(\rho,X)$ in such a state has a good extrapolated limit and it makes sense to write
\be
\lim_{\rho \rightarrow \infty} D(\rho) \langle \Omega | \phi(\rho, X) | \Omega \rangle  = \langle \Omega| \opspat(X) | \Omega \rangle.
\ee
Since $|\Omega \rangle$ has bounded $\nu$-content the limit exists in spite of the growing exponentials in \eqref{extrapolosc}.

Moreover, in free field theory the expectation value of a normal ordered product in such a state also has a good extrapolated limit. Therefore one can write
\be
\label{compactnustate}
\lim_{\rho_{i}\, \rightarrow\, \infty} \, \left(\prod_i D(\rho_i) \right)  \, \langle\Omega\vert\, :\phi(\rho_{1}, X_{1})\, \dots, \phi(\rho_{n}, X_{n}):\, \vert\, \Omega\, \rangle\, =\, \langle \Omega | :\opspat( X_1) \ldots \opspat( X_n):| \Omega \rangle. 
\ee

The notion of normal ordering is not especially useful in the interacting theory. Moreover, products  of normal ordered operators are not normal ordered and so the expectation value of products of normal ordered field insertions might not have a good limit at $\spatinf$. Therefore restricting to normal-ordered operators does not define a satisfactory algebra of observables at $\spatinf$.  We will show below that smearing the extrapolated operators provides a more satisfactory route to obtaining an well-behaved operators.

\subsection{Smeared boundary operators \label{subsecsmeared}}
 We now show that by smearing boundary operators with an appropriate class of smearing functions, one obtains observables with finite fluctuations. We show that this set of  class of operators is rich enough to encode all local bulk operators.
We consider the limit of smeared field operators,
\be
\label{smearedphi}
\opspat( \gfull) \equiv \int \opspat( X) \gfull(X) \dds, 
\ee
for a specific class of functions $\gfull(X)$ that we now detail. Here, and in what follows $\dds$ is the invariant measure on de Sitter space that in the $(\tau, \Omega)$ coordinates takes the form
\be
\dds = \cosh^2 \tau d \tau d \Omega.
\ee

To determine the allowed class of functions, we demand that the fluctuations of the boundary operator in the vacuum, $\langle 0 |\opspat( \gfull)^2 |0 \rangle$, be finite. The allowed class of functions can be understood quite easily if decomposed using the functions $\orthfunc_{\nu,\ell}$, defined in \eqref{orthfunc}. Without loss of generality, we consider real functions $g(X)$ so that $\opspat(g)$ is Hermitian. The space of allowed functions that we find below will form a complex vector space, and so complex linear combinations of these real smearing functions will also yield operators with finite fluctuations. We consider functions of the form
\be\label{gnulpsibasis}
\gfull(X) = \sum_{\vec{\ell}} \int_{0}^{\infty} d \nu \,  \gfull_{\nu, \vec{\ell\,}}  \, \orthfunc_{\nu,\ell}(\tau) \, Y^*_{\vec{\ell}}(\Omega)\, +\,  \hc
\ee

Recall the mode expansion of the extrapolate operator \eqref{extrapolosc}. The benefit of using the decomposition of the smearing function above is that we can now immediately perform the integral over de Sitter space and use the orthogonality relations in \eqref{orthognalityfuncorthfunc} to get
\be
\label{gfullexpansion}
\opspat(\gfull) = \sum_{\vec{\ell\,}} \int d \nu \tN_{\nu} \, \left(  \tB_{\nu, \vec{\ell\,}} \, \gfull_{\nu, \vec{\ell \, }} +    \tB^{\dagger}_{\nu, \vec{\ell\,}} \, \gfull_{\nu, \vec{\ell \, }^*}  \right). 
\ee
The fluctuations of $Z(\gfull)$ in the vacuum are given by
\be\label{ffgnul}
\langle 0 | \opspat(\gfull) \opspat(\gfull) | 0 \rangle = \sum_{\vec{\ell\,}} \int_0^{\infty} d \nu |\tN_{\nu}|^2 |\gfull_{\nu, \vec{\ell}}|^2. 
\ee
Given that the normalization factor (in \eqref{normeuclidean}) grows as  $\tN_{\nu}\propto e^{\pi \nu\over 2}$,  we see that, as long as $\gfull_{\nu, \ell}$ is nonsingular,  $Z(\gfull)$ has finite fluctuations in the vacuum provided the following two conditions are met
	\be\label{smearingnufalloff}
		\begin{split}
		\gfull_{\nu,\vec{\ell\,}} \ &\xrightarrow[\nu \rightarrow \pm \infty]{} \  \, \frac{e^{- \frac{\pi|\nu|}{2}}}{\nu^{\alpha}} , \qquad \alpha\, >\, \frac{1}{2}\, \forall\, \vec{\ell},\\
		\sum_{m} |\gfull_{\nu,\vec{\ell\,}}|^2 \ \ &\xrightarrow[\ell \rightarrow \infty]{}  \  \ \ \, {1\over {\ell}^{\beta} }, \qquad \qquad \beta > 1 \,  \forall \, \nu.
		\end{split}
	\ee

If we consider two smearing functions, $\gfull_1$ and $\gfull_2$, both of which have the decay above, then a similar argument shows that $\langle 0 | \opspat(\gfull_1) \opspat(\gfull_2) | 0 \rangle$ is also finite. Since, in the absence of interactions, all $n$-point correlation functions can be decomposed into products of two point function, this class of smeared operators defines an algebra of boundary operators with finite fluctuations.  Therefore we conclude that elements of the algebra of observables
	\be
		\begin{split}
		\alset_{\spatinf} = \text{span}\{ \opspat(\gfull^1) \ldots \opspat(\gfull^n)  \},
		\end{split}
	\ee
	where the smearing functions $\gfull^i$ are nonsingular, and their decomposition in terms of \eqref{gnulpsibasis} satisfies the property \eqref{smearingnufalloff}, have finite fluctuations.

We conclude with two comments:
\begin{enumerate}
\item
First, note that the conditions on the state $|\Omega \rangle$ that appears in \eqref{compactnustate} can be likewise weakened. The equation \eqref{compactnustate} holds provide the $\nu$-content and $\ell$-content of $|\Omega \rangle$ dies off sufficiently fast even if it is not strictly compact.
\item
Second, it is sometimes convenient, in intermediate calculations, to consider smearing functions that violate \eqref{smearingnufalloff}. For instance, when we reconstruct bulk modes we will study $g_{\nu, \ell}$ with delta-function support. Similarly, when we reconstruct bulk local operators in terms of boundary operators we will find smearing functions where the large $\nu$ falloff corresponds to \eqref{smearingnufalloff} but with $\alpha = {1 \over 2}$ or $\beta = 1$. Such operators do not have finite fluctuations for the simple reason that a local bulk operator or an infinitely-sharp mode in momentum space does not have finite fluctuations. But these divergences --- and their cure which is to simply smear the operators slightly, over and above the boundary smearing described above --- are familiar from the study of quantum field theory.
\end{enumerate}

\subsubsection*{Relation with the standard QFT algebra}
 We now show that, in free-field theory,  any bulk local operator  can be written as an appropriately smeared boundary operator.

 This may appear confusing at first sight, since boundary operators appear to be spacelike separated from every bulk operator. 
However, as explained below \eqref{blowupmetric}, the $\dsthree$ slice at any finite value of $\rho$ captures the entire exterior of the light cone. Therefore if one considers smearing functions with suitably tuned behaviour at $\tau \rightarrow \pm \infty$ and then takes the $\rho \rightarrow \infty$ limit, it is still possible to recover operators everywhere on a spacelike slice in the bulk from $\spatinf$.

This result is easiest to see if we work in the spherical basis. We first note that boundary operators can be used to reconstruct the mode operators $\tB_{\nu,\vec{\ell \, }}$ and $\tB_{\nu,\vec{\ell \, }}^{\dagger}$. From the expansion of extrapolate operator \eqref{extrapolosc} and the orthogonality relations \eqref{orthognalityfuncorthfunc}, we find that
\be\label{smearingosc}
	\begin{split}
		\tB_{\nu,\vec{\ell \, }} &=  \int d\tau \, d\sph \, \cosh^2(\tau) \, \opspat(\tau, \sph) \, \left[ {1\over \tN_{\nu} }  \, \orthfunc_{\nu, \ell} (\tau)  \, Y^*_{\vec{\ell\,}}(\sph)   \right],\\
		\tB^{\dagger}_{\nu,\vec{\ell \, }} &=  \int d\tau \, d\sph \, \cosh^2(\tau) \, \opspat(\tau, \sph) \, \left[ {1\over \tN_{\nu} }  \, \orthfunc^*_{\nu, \ell} (\tau)  \, Y_{\vec{\ell\,}}(\sph)   \right].
	\end{split}
\ee
Note that when expanded using \eqref{gnulpsibasis} the smearing function ${1\over \tN_{\nu} }  \, \orthfunc_{\nu, \ell} (\tau)  \, Y^*_{\vec{\ell\,}}(\sph)$  has  $\delta$-fn support in $(\nu,\, \ell)$ so it satisfies the falloff \eqref{smearingnufalloff}, albeit with delta-function supported $\gfull_{\nu, \ell}$. But, in the free theory, the entire algebra of massive field operators is generated by polynomials of $\tB_{\nu, \vl}$ and $\tB^{\dagger}_{\nu, \vl}$. Therefore polynomials of appropriate smeared asymptotic operators suffice to generate the entire algebra.

\subsection{Decay of the smearing function and localization of bulk fields \label{subsecdecay}}

Even though all operators on a Cauchy slice can be recovered by smearing boundary operators, we now show that a sense of operators localized at ``large $\rho$'' is obtained by smearing operators on $\spatinf$ with rapidly decaying smearing functions. The intuitive idea is that smearing functions that are, in an appropriate sense, ``concentrated'' near $\tau = 0$ lead to bulk operators that are localized at large $\rho$. More precisely,  we will prove the following result.  Consider a smearing function $g(\tau, \Omega)$ on $\dsthree$ with the property that, as $\tau \rightarrow \infty$, we have
\be
\label{smearingcond}
\lim_{\tau \rightarrow \infty} g(\tau, \Omega) e^{\delta^{-1} \cosh \tau} = 0.
\ee
This means that $g(\tau, \Omega)$ decays faster than $e^{-\delta^{-1} \cosh \tau}$ along every direction in $\Omega$.  Then we will show that
\be
\label{localitycondition}
\forall \rho_0 < {1 \over 2m \delta} \quad \text{we~have~} \quad  [\opspat(g) , \phi(\rho_0, \tau' = 0, \Omega')] = 0.
\ee
This means that the operator $\opspat(g)$ has no overlap with the algebra of operators localized in the region with $\rho < \rho_0$ on the zero-time slice.

To prove the result, we start with the representation of the boundary operator as a limit of the bulk operator and consider the commutator
\be
\label{commutatorlargerho}
[\opspat(g) , \phi(\rho_0, \tau' = 0, \Omega')] =  \lim_{\rho \rightarrow \infty} \int \cosh^2 \tau d \tau d \Omega \, D(\rho) \int [\phi(\rho, \tau, \Omega) g(\tau, \Omega), \phi(\rho_0, \tau' = 0, \Omega')] . 
\ee
By bulk microcausality (recall that we are studying nongravitational theories  and so microcausality holds as an exact statement), the commutator in the integrand can only receive contributions from those values of $\tau, \Omega$ that satisfy
\be
|\rho (\sinh \tau, \cosh \tau \hat{\Omega}) - \rho_0 (0, \hat{\Omega}')|^2 \leq 0,
\ee 
i.e. when the bulk point and the point near the boundary are separated by either a timelike or a null interval.  This condition can be rewritten as
\be
\rho^2 + \rho_0^2 - 2 \cosh \tau \rho \rho_0 \hat{\Omega} \cdot \hat{\Omega}' \leq 0.
\ee
Noting that $\hat{\Omega} \cdot \hat{\Omega}' \leq 1$ we see that for any nonzero $\rho_0$ this condition requires
\be
\label{largetaucondition}
\cosh \tau >  {1 \over 2} {\rho \over \rho_0}. 
\ee

\begin{figure}[!h]
\begin{center}
\begin{subfigure}{0.45\textwidth}
\begin{center}
\includegraphics[width=\textwidth]{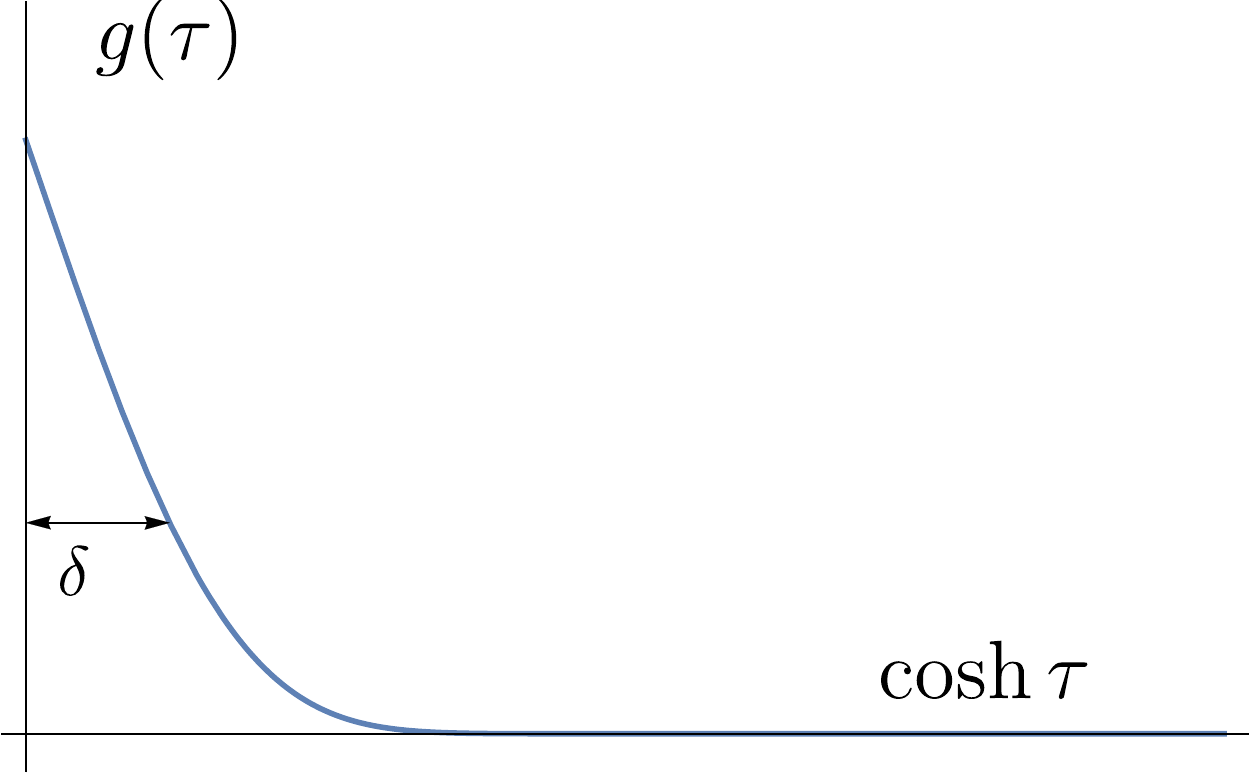}
\end{center}
\end{subfigure}
\begin{subfigure}{0.45\textwidth}
\begin{center}
\includegraphics[height=0.25\textheight]{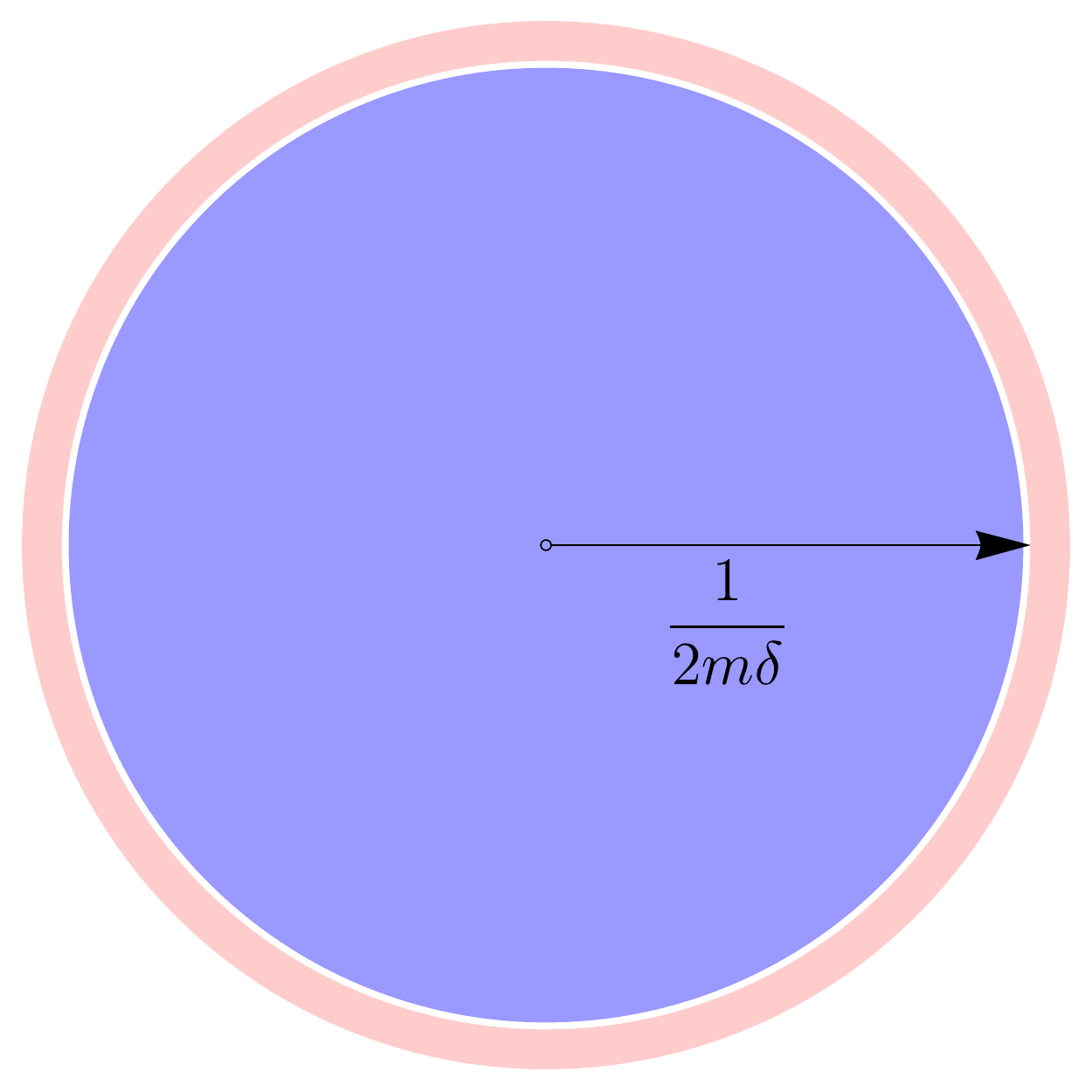}
\end{center}
\end{subfigure}
\caption{\em A ``narrow'' smearing function leads to operators in the exterior of a large ball in the bulk. More precisely, the operator obtained by smearing $\opspat(\tau, \Omega)$  with a function that decays asymptotically like $e^{-\delta^{-1} \cosh \tau}$ (shown in the left subfigure) commutes with all operators in a ball of radius ${1 \over 2 m \delta}$ (shown in blue on the right). This means that such operators are necessarily localized in the region outside this ball (shown in pink).  \label{fignarrowsmearing}}
\end{center}
\end{figure}

Therefore as we take $\rho \rightarrow \infty$, the commutator receives contributions from larger and larger values of $\tau$ with the lower bound given by \eqref{largetaucondition}. In this limit, when $\tau$ is increased with $\rho$ to meet \eqref{largetaucondition} and when the smearing functions satisfies \eqref{smearingcond} then 
\be
\forall \rho_0 < {1 \over 2 m \delta}, \quad \text{we~have} \quad \lim_{\rho \rightarrow \infty} D(\rho) g(\tau, \Omega) = 0. 
\ee
Since the commutator on the right hand side of \eqref{commutatorlargerho} is not expected to grow with $\rho$ we see that the limit on the right hand side of \eqref{commutatorlargerho} vanishes. The result \eqref{localitycondition} follows immediately. Figure \ref{fignarrowsmearing} depicts this result graphically.

\subsection{Transformation of asymptotic operators under global translations \label{subsectransform}}
Next, we investigate the transformation properties of asymptotic operators under global spacetime translations. The coordinate system that we are using breaks translational invariance. Nevertheless we show that, at least for a certain range of parameters, boundary operators transform in a simple manner under translations.

Consider a transformation by a four-vector $y$ which is implemented by the unitary operator $e^{i P \cdot y}$ where $P \equiv (H, \vec{P})$ is the four-vector operator comprising the Hamiltonian and the momentum. We find that
\be
e^{-i P \cdot y} \phi(\rho, X) e^{i P \cdot y}  = \phi(\rho', X'), 
\ee
where, at large $\rho$,
\be
\begin{split}
	&\rho' = \sqrt{(\rho X + y)^{2}} = \rho + X \cdot y + \Or[{1 \over \rho}]; \\
	&X' = {\rho X + y \over \rho'} = X + \Or[{1 \over \rho}].
\end{split}
\ee
We also note, that with $\rho'$ given by the expression above,
\be\label{drhodrhoprime}
D(\rho') = D(\rho) e^{m X \cdot y} \left(1 + \Or[{1 \over \rho}] \right). 
\ee

Combining these equations we find that
\be
\label{pointwisetrans}
\langle \Psi_1 | e^{-i P \cdot y} \opspat(X) e^{i P \cdot y} | \Psi_2 \rangle  =  e^{-m y \cdot X} \langle \Psi_1 |   \opspat(X) | \Psi_2 \rangle,
\ee
{\em subject to the condition} that the external states $|\Psi_1 \rangle$ and $|\Psi_2 \rangle$ are such that the insertion of the pointwise operator $\opspat(X)$ makes sense. This caveat is necessary since the result above assumes that the large-$\rho$ limit exists both for the insertion of $D(\rho) \phi(\rho, X)$ and for the insertion of $D(\rho') \phi(\rho', X')$.   

For a smeared operator, we have
\be
\label{smearedtrans}
\int \langle \Psi_1 | e^{-i P \cdot y} \opspat(X) \gfull(X) e^{i P \cdot y} | \Psi_2 \rangle \dds  = \int \dds  e^{-m y \cdot X} \langle \Psi_1 |  \gfull(X) \opspat(X) | \Psi_2 \rangle.
\ee
Therefore a spacetime translation with spacetime vector $y$ can be thought of as modifying the smearing function from $\gfull(X) \rightarrow e^{-m y \cdot X} \gfull (X)$. 

We note that, for some purposes, it may be useful  to consider translations with respect to imaginary spacetime shifts.
\be
\label{imagtimeshift}
\int \langle \Psi_1 | e^{P \cdot y} \opspat(X) \gfull(X) e^{-P \cdot y} | \Psi_2 \rangle \dds  = \int \dds  e^{-i m y \cdot X} \langle \Psi_1 |  \gfull(X) \opspat(X) | \Psi_2 \rangle.
\ee

\paragraph{Constraints on asymptotic time translations.}
Caution must also be exercised in implementing \eqref{smearedtrans} since the factor $e^{-m y \cdot X}$ grows doubly exponentially for large values of the  $\tau$-coordinate. Therefore, it must be ensured that the smearing function dies off rapidly enough in the first place in order for the transformed smearing function to still have well-defined expansion coefficients \eqref{gfullexpansion}. If the original smearing function $\gfull(X)$ does not die off rapidly and  $e^{-m y \cdot X} \gfull(X)$ grows at large positive or negative $\tau$ then \eqref{smearedtrans} cannot be used. This constraint must be kept in mind to avoid obtaining contradictory results by the application of \eqref{smearedtrans}. We will see an explicit example of this constraint in the next subsection.

We note that imaginary translations displayed in \eqref{imagtimeshift}  act more smoothly on the smearing function since the additional factor no longer grows sharply in magnitude at large $\tau$, although it oscillates rapidly.

\subsection{Boundary representation of local bulk operators \label{brobo}}
To conclude this section, we derive the expression for a bulk quantum-field operator in terms of operators at $\spatinf$. This means that we write a bulk operator as 
\be\label{bbhkllf}
	\phi(\rho_0,X_0) = \int \dds \, \smrf(\rho_0,X_0|X) \, \opspat(X),
\ee
using a smearing function, $\smrf$. An  analogous construction is well known in the context of AdS/CFT as the HKLL construction \cite{Hamilton:2007wj,Hamilton:2006fh,Hamilton:2006az,Hamilton:2005ju}.

Let us recall the bulk quantum field expansion in the spherical basis,
\be
\phi(\rho_0,X_0)=\frac{1}{\rho_0}\int d\nu \,\tN_{\nu} \sum_{\vec{\ell}} {K_{i \nu}(m \rho_0) \over \rho_0} \tB_{\nu, \vec{\ell}} \,\func_{\nu,\ell}(\tau_0) Y_{\vec{\ell}}\,(\Omega_0) + \hc,
\ee
where, as usual, the bulk coordinates are given by $(\rho_0,X_0)=(\rho_0,\rho_0 \sinh \tau_0,\rho_0 \cosh \tau_0 \, \hat{\Omega}_0)$. Using the boundary representation of oscillators in \eqref{smearingosc}, we immediately obtain
\be
	\phi(\rho_0,X_0)=\int d\tau d\sph \cosh^2 \tau \, \opspat(\tau,\sph)\int d\nu \, \sum_{\vec{\ell}} {K_{i \nu}(m \rho_0) \over \rho_0} \left[ \orthfunc_{\nu, \ell} (\tau)  \, Y^*_{\vec{\ell\,}}(\sph)   \right] \,\func_{\nu,\ell}(\tau_0) Y_{\vec{\ell}}\,(\Omega_0) + \hc
\ee
Comparing with \eqref{bbhkllf}, we get
\be\label{localsmearing}
	\smrf(\rho_0,\tau_0 ,\sph_0|\tau,\sph) = \sum_{\vec{\ell}} \int d\nu {K_{i \nu}(m \rho_0) \over \rho_0} \orthfunc_{\nu, \ell} (\tau)  \, Y^*_{\vec{\ell\,}}(\sph)  \,\func_{\nu,\ell}(\tau_0) Y_{\vec{\ell}}\,(\Omega_0) + \hc
\ee

We now verify that when this smearing function $\smrf$ is decomposed in coefficients,  $\smrf_{\nu,\vec{\ell\,}}$ ,  as displayed in \eqref{gnulpsibasis}, those coefficients satisfy the large-$\nu$ decay shown in \eqref{smearingnufalloff} albeit with $\alpha = {1 \over 2}$ and $\beta = 1$ as mentioned above. Comparing with \eqref{gnulpsibasis}, one can directly read off $\smrf_{\nu,\vec{\ell\,}}$ from \eqref{localsmearing}.  
\be
\smrf_{\nu,\vec{\ell\,}}(\rho_0,\tau_0 ,\sph_0) =  \frac{1}{\rho_{0}}\, {K_{i\nu}(m\rho_{0})} \, \func_{\nu, \ell}(\tau_{0}) \, Y_{\vec{\ell\,}}(\sph_0).
\ee
Since $\func_{\nu, \ell}$ is $O(\nu^0)$ at large $\nu$, we explicitly see that,
\be
\begin{split}
&\smrf_{\nu,\vec{\ell\,}}(\rho_0,\tau_0 ,\sph_0)\, \xrightarrow{\nu\, \rightarrow\, \infty}\, \frac{1}{\sqrt{\nu}}\, e^{-\frac{\pi\nu}{2}};\\
&\sum_{m} |\smrf_{\nu,\vec{\ell\,}}(\rho_0,\tau_0 ,\sph_0)|^2\, \xrightarrow{\ell \, \rightarrow\, \infty}\, {1 \over \ell};
\end{split}
\ee
at large $\nu$ and $\ell$  using \eqref{largenulimit}. As expected, this satisfies \eqref{smearingnufalloff} with $\alpha = {1 \over 2}$ and $\beta = 1$. An operator with finite fluctuations can be obtained by slightly smearing the bulk operator.

\subsubsection*{Example: s-wave}
To gain a better understanding of the smearing function, let us evaluate it explicitly in a simple case.  The simplest example is where we consider the s-wave component of the bulk field, i.e., we consider the bulk operator obtained by integrating the bulk field on a sphere. 
\be\label{smearingexample}
\phi_{\ell = 0}(\rho_0,\tau_0) \defeq \int d\sph_0 \, \, \phi(\rho_0, \tau_0, \sph_0) . 
\ee
Clearly, the smearing function that produces this operator is the following.
\be
	\smrf_{\ell = 0} (\rho_0,\tau_0|\tau) = \int d\sph_0  \, \smrf(\rho_0,\tau_0 ,\sph_0|\tau,\sph). 
\ee
Note that the left hand side only depends on $\tau$ because after doing the integral over $d \sph_0$ the dependence on the coordinate $\sph$ will drop out.
Using \eqref{localsmearing}, we get,
\be
	\smrf_{\ell = 0} (\rho_0,\tau_0|\tau)  =  {1\over \rho_0} \int\limits_0^{\infty} d\nu K_{i\nu}(m\rho_0) \, \func_{\nu, \ell=0}(\tau_0)\, \orthfunc_{\nu,\ell=0}(\tau) + \hc
\ee
The spherical mode functions for $\ell = 0$ take on the simple form
\be
\begin{split}
&\func_{\nu, \ell=0}(\tau_0) = \left(e^{-i \nu \tau_0} - e^{-\pi \nu} e^{i \nu \tau_0} \right) \text{sech} \tau_0; \\
&\orthfunc_{\nu, \ell=0}(\tau) = \frac{e^{\frac{\pi  \nu }{2}} \text{csch}(\pi  \nu ) \cosh \left(\frac{1}{2} \nu 
   (\pi +2 i \tau )\right)}{2 \pi } \text{sech}(\tau).
\end{split}
\ee
A little algebra then yields the expression
\be
\smrf_{\ell = 0} (\rho_0,\tau_0|\tau) = {1\over \pi \rho_0 \cosh \tau_0 \cosh \tau } \int\limits_0^{\infty} d\nu K_{i\nu}(m\rho_0) \, \cos(\nu \, (\tau - \tau_0)) . 
\ee
Using the identity (see chapter 12 of \cite{Erdelyi:1954:TI2}),
\be
\int_{0}^{\infty} d\nu \, \cos( \nu \tau) \, K_{i \nu}(m \rho_0) = {\pi \over 2} e^{-m\rho_0 \cosh(\tau)},
\ee
we arrive at a simple form for the smearing function
\be\label{swavesmearing}
\smrf_{\ell = 0} (\rho_0,\tau_0|\tau) ={1 \over 2}\frac{1}{\rho_0 \cosh{\tau_0} \cosh \tau} e^{-m\rho_0 \cosh(\tau_0-\taub)}. 
\ee

We can use the formula above to verify our expression for the transformation of the smearing function under time translations. Given a point $(\rho_0, \tau_0 , \Omega)$ in the space-like wedge, a global time translation $t\, \rightarrow\, t + \eta$  maps it to a point in the space-like wedge only if $(\rho_0 \sinh (\tau_0) + \eta)^2 < \rho_0^2 \cosh^2 \tau_0$ with new coordinates 
\be
\rho_0' = \left(\rho_0^2 \cosh^2 \tau_0 - (\eta + \rho_0 \sinh \tau_0)^2  \right)^{1 \over 2}; \qquad \tau_0' = \tanh^{-1} \left(\rho_0 \sinh \tau_0 + \eta  \over \rho_0 \cosh \tau_0 \right). 
\ee
Within this range, the formula \eqref{smearedtrans} can be applied and yields a result consistent with \eqref{swavesmearing} since 
\be
\label{newsmearf}
\begin{split}
\smrf_{\ell=0}(\rho_0', \tau_0'|\tau) &= {1 \over 2}\frac{1}{\rho_0' \cosh{\tau_0'} \cosh \tau} e^{-m\rho_0' \cosh(\tau_0'-\taub)} \\
&={1 \over 2}\frac{1}{\rho_0 \cosh{\tau_0} \cosh \tau} e^{-m\rho_0 \cosh(\tau_0-\taub) + m \eta \sinh \tau}.
\end{split}
\ee
However, it is not permissible to use the formula \eqref{smearedtrans} for generic time translations since, as explained above,  when the condition on $\eta$ above is not met, the new smearing function \eqref{newsmearf} does not decay at large $\tau$.

\section{Extrapolate dictionary and boundary observables for  interacting fields \label{s_interactions}}
In this section, we argue that the extrapolate dictionary continues to be valid even in the interacting theory, subject to the same infrared subtleties that we described in the free theory. We also explain which results from the free theory generalize to the interacting theory and mention some additional subtleties that arise in the interacting case. In this section, we will provide a quick  analysis utilizing Feynman diagrams.  In the next section, we provide a more detailed analysis of the interacting theory using the asymptotic form of the equations of motion.

\subsection{Validity of the extrapolate dictionary}

Consider the insertion of a Minkowski field at large $\rho$ inside a correlation function in the interacting theory of the form $\langle \ldots \phi(\rho, X) \ldots \rangle$. We wish to take the large $\rho$ limit of this correlator to obtain observables at $\spatinf$. Since our interest is in defining an algebra of observables at $\spatinf$, it is natural to consider the expectation value of the product of such operators. The expectation value of a product of operators corresponds to a Wightman correlator, as mentioned previously.

For a quick review of perturbation theory for Wightman correlators, we refer the reader to appendix A.2 of \cite{Banerjee:2019kjh}.\footnote{The Feynman rules provided there generalize immediately to the case of interest here except that one must set $\beta \rightarrow \infty$, which leads to a replacement of thermal propagators by vacuum propagators.} Although textbook discussions often restrict attention to time-ordered correlators, the standard interaction-picture perturbative formalism can be used to easily derive Feynman rules for Wightman correlators.  Here, we need only one salient feature of these Feynman rules. Wightman perturbation theory requires the use of four kinds of propagators: (i) the standard time-ordered propagator $\langle 0 | {\cal T}\{\phi(\rho, X_1) \phi(\rho, X_2) \} | 0 \rangle$ (ii) the anti-time-ordered propagator $\langle 0 | \overline{{\cal T}}\{\phi(\rho, X_1) \phi(\rho, X_2) \} | 0 \rangle$ (iii) the Wightman function $\langle 0 | \phi(\rho, X_1) \phi(\rho, X_2) | 0 \rangle$ (iv) the conjugate of the Wightman function  $\langle 0 | \phi(\rho, X_2) \phi(\rho, X_1) | 0 \rangle$.

\begin{figure}[!h]
\begin{center}
\includegraphics[height=0.3\textheight]{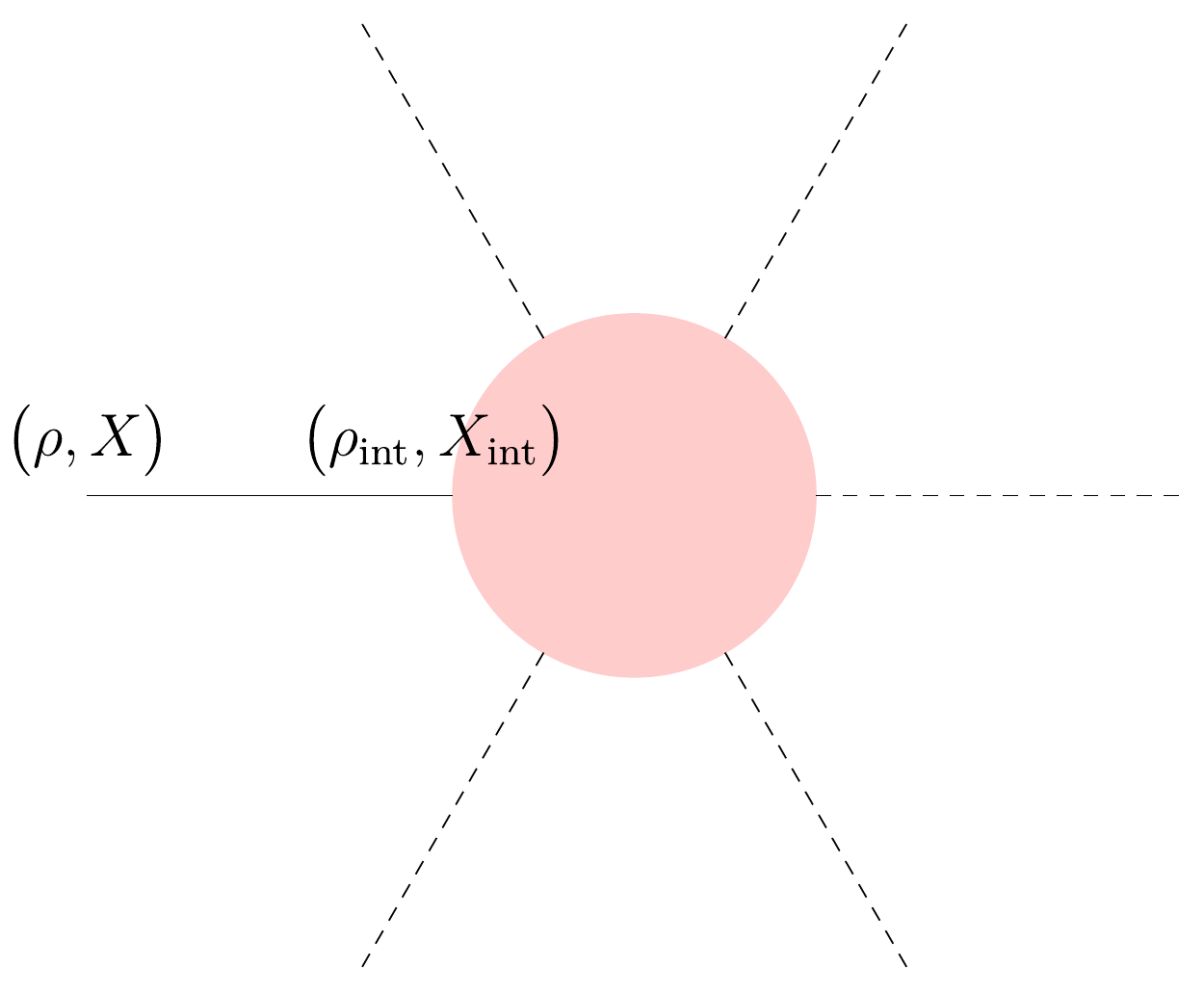}
\caption{\em A schematic Feynman diagram showing the contribution of interactions. We are interested in the propagator that connects the external point to an internal point. The dashed lines indicate other possible external legs and the blob may include an arbitrary subdiagram. \label{figsamplefeyn}}
\end{center}
\end{figure}

Although different kinds of propagators appear, a Wightman correlator with an insertion of $\phi(\rho, X)$  is still  computed by a sum of Feynman diagrams of the form shown in \ref{figsamplefeyn}. Each Feynman diagram contains a propagator that connects the point $(\rho, X)$ to the rest of the diagram. This leads to 
\be
\label{generalexpression}
\langle \ldots \phi(\rho, X) \ldots \rangle = \int H(\rho_{\text{int}}, Y_{\text{int}}) G(\rho_{\text{int}}, Y_{\text{int}}, \rho, X) d Y_{\text{int}} d \rho_{\text{int}} , 
\ee
where $G(\rho_{\text{int}}, Y_{\text{int}}, \rho, X)$ is one of the four possible propagators listed above and  $H(\rho_{\text{int}}, Y_{\text{int}})$ contains the contribution from the rest of the Feynman diagram. The expression above assumes that the external point is connected to an internal point but the argument below generalizes trivially to the case where it is two external points that are connected.  (The reader who prefers momentum space can consult a parallel analysis in section \ref{secasympanal}.)

In perturbation theory, one must account for self-energy loops in the external propagators. However, we expect, on general grounds, that the renormalized propagator will have a Kallen-Lehmann spectral representation. This means that
\be
G(\rho_{\text{int}}, Y_{\text{int}}, \rho, X) = \int_0^{\infty} d s \,  \mu(s) G_s(\rho_{\text{int}}, Y_{\text{int}}, \rho, X), 
\ee
where $\mu(s)$ is some positive weight function and $G_s$ is the free propagator for a field of mass $s$ that we have encountered in the previous section. This free-field propagator is of the form
\be
G_s(\rho_{\text{int}}, Y_{\text{int}}, \rho, X) = {-i \over 2 \pi^3} \int_0^{\infty} \nu d \nu {K_{i \nu}(m \rho) K_{i \nu}(m \rho_{\text{int}}) \over \rho \rho_{\text{int}}} C^{1}_{i \nu - 1}(-X \cdot Y_{\text{int}}). 
\ee
As explained below \eqref{twopoint}, this form accounts for all four possible propagators listed above provided we use different $i \epsilon$ prescriptions for the expression $-X \cdot Y_{\text{int}}$.  

We now assume that, for stable massive particles,  the spectral function $\mu(s)$ has a delta function at $s = m$ where $m$ is the renormalized mass of the field plus additional contributions for higher values of $s$ 
\be
\mu(s) = \delta(s - m) + \ldots
\ee
and does not have support below the mass of the particle, 
\be
\mu(s)  = 0, ~~\text{if}~~ s < m.
\ee

If this assumption is satisfied then we see that the expression \eqref{generalexpression} has a good extrapolated limit. This is because each Macdonald function that appears above has a large-$\rho$ falloff that depends on $s$ but the {\em slowest} falloff is determined by the smallest value of $s$ for which $\mu(s)$ has support. In the limit as $\rho \rightarrow \infty$ it is only this slowest falloff that is important. Therefore
\be
\label{greenextrapolate}
\lim_{\rho \rightarrow \infty} D(\rho) \int ds^2 \mu(s^2) G_s(\rho, X, \rho_{\text{int}}, Y_{\text{int}}) = \int_0^{\infty} \nu d \nu K_{i \nu}(m \rho_{\text{int}})  C^{1}_{i \nu - 1}(-X \cdot Y_{\text{int}}). 
\ee

Since $G(\rho_{\text{int}}, Y_{\text{int}}, \rho, X)$ in \eqref{generalexpression} has a good extrapolated limit and it is the only term in \eqref{generalexpression} that depends on $\rho$, this suggests that the Heisenberg operator itself has a good extrapolated limit. 

Therefore our argument suggests that in the interacting theory, massive field operators continue to have a falloff at the rate $D(\rho)^{-1}$. This is the same as the free theory except that, as one would expect, it is the  renormalized mass of the field that enters this expression.

\paragraph{Smearing functions in the interacting theory. \label{interactsubtlety}}
The integrand in \eqref{greenextrapolate} has the same growth at large $\nu$ that was discussed in the free-field case in section \ref{subsecsubtlety}. This is because when its argument is negative, the Gegenbauer function grows as $e^{\pi \nu}$ at large $\nu$ (as can be seen from \eqref{gegenelem}) whereas the Macdonald function only decays as $e^{-\pi \nu \over 2}$ as shown in \eqref{largenulimit}. 

Therefore it is clear that, just as in the free-field case, the expression \eqref{greenextrapolate} must be interpreted as a distribution that makes sense {\em after} it is integrated with an appropriate smearing function that depends on  $X$ and the coordinates of  the other external legs in the diagram \eqref{figsamplefeyn}.

An important question that we do not settle in this paper is about the nature of allowed smearing functions in the interacting theory. Phrased alternately, since we must think of correlators of interacting fields as distribution, what is the domain of functions that these distributions act upon?  Is the domain the same as the domain found in the free theory and discussed in section \ref{subsecsmeared} or is it distinct? It is possible that interacting correlators make sense only after smearing with an appropriate ``joint-smearing'' function rather than when each external point is smeared independently. We leave further investigation of this issue to future work.

\subsection{Properties of the interacting theory}
We now briefly list those properties of the free theory that do not rely on the free mode expansion and therefore carry over to the interacting case.

\begin{enumerate}
\item
The results of section \ref{subsecdecay} remain true in the interacting theory. This means that a smeared asymptotic operator $\opspat(g)$ where the smearing function dies off faster than $e^{-\delta^{-1} \cosh \tau}$  commutes with all bulk operators localized in the ball with $\rho < {1 \over 2 m \delta}$ on the zero-time slice. This is because the results of section \ref{subsecdecay} only rely on microcausality, which is expected to be exact in any nongravitational theory.

We mentioned the possibility that the class of allowed smearing functions might need to be refined in the interacting theory and perhaps only a class of jointly smeared operators is admissible. The results of section \ref{subsecdecay} still hold in the following form. Consider the operator 
\be
\label{jointlysmeared}
\int  \left( \prod_{i}\cosh^2 \tau_i d \tau_i d \Omega_i \right) \opspat(\tau_1, \Omega_1) \opspat(\tau_2, \Omega_2) \ldots \opspat(\tau_n, \Omega_n) g(\tau_i, \Omega_i), 
\ee
where the product of operators is smeared with a joint smearing function. If $g$ has the property that it dies off faster than $e^{-\delta^{-1} \cosh \tau_i}$ for each $\tau_i$ then the jointly smeared operator still commutes with all bulk operators localized in the region with $\rho < {1 \over 2 m \delta}$ on the zero-time slice.

It appears clear that the results of section \ref{subsecdecay} can be generalized to nongravitational interacting theories in a spacetime that is only asymptotically flat. QFTs in curved spacetime also satisfy microcausality \cite{Dubovsky:2007ac}. So, in such theories asymptotic operators smeared with doubly-exponentially decaying smearing functions commute with bulk operators inside a region, although the precise relationship between the rate of decay of the smearing function and the boundary of the bulk region depends on the details of the background metric. 

In gravitational theories, the results of section \ref{subsecdecay} do not hold. On the other hand, it is still sensible to consider asymptotic operators smeared with rapidly decaying smearing functions. One can then study how the properties of this asymptotic algebra differ between gravitational and nongravitational theories. This will be discussed at greater length in \cite{Laddha:2022masshol}.
\item
The results of section \ref{subsectransform} on the transformation of asymptotic operators under Poincare transformations are expected to hold in the interacting theory. These results only invoke the action of large diffeomorphisms in the asymptotic part of the spacetime,  which is independent of the details of the interaction. Again, considering the jointly smeared operator \eqref{jointlysmeared}, under a Poincare transformation we expect that under a translation with four-vector $y$, the smearing function transforms as
\be
g(X_i) \rightarrow e^{-m y \cdot \sum_{k} X_{k}} g(X_i). 
\ee
This formula, of course, assumes that both sides of the equation above define valid smearing functions.
The transformation of asymptotic operators is expected to be correct even for gravitational theories in asymptotically flat space, since it does not rely on an assumption of microcausality.
\end{enumerate}

\subsection{Differences between the interacting and free theories \label{diffintfree}}
On the other hand, there are also some important differences between the interacting and free theory.
\begin{enumerate}
\item
The mode expansion used in the free-field theory, as for example, in \eqref{gfullexpansion} does not hold  in the interacting theory.
\item
Relatedly, correlators of asymptotic operators do not factorize into products of two-point functions. For instance, with a $\phi^3$ interaction, as we argue in the next section, we expect that  the correlator of three smeared boundary operators does not vanish. In this sense, $\spatinf$ differs from ${\cal I}^{\pm}$ and $i^{\pm}$ where the theory becomes free. But $\spatinf$ is similar to the boundary of AdS where interactions persist even asymptotically.
A corollary of this is that operators at $\spatinf$ cannot be identified with operators at $i^{+}$ or operators at $i^{-}$ in the interacting theory.
\item
The map between boundary and bulk operators detailed in section \ref{brobo} must be corrected in the interacting theory. Such corrections to the bulk-boundary map have been explored in the context of AdS/CFT \cite{Kabat:2012av}. We leave investigation of such corrections to future work. 
\end{enumerate}

\section{More on interactions  \label{secasympanal}}
In previous sections, we have studied an extrapolate dictionary in the language of quantum field theory and Feynman diagrams. In this section, we will present a complementary perspective on this dictionary.

First we rederive the extrapolate dictionary using the  classical equations of motion. Conventionally, this has been the approach used in studies of asymptotic quantization both at ${\cal I}^{\pm}$ and at $i^{\pm}$ \cite{Campiglia:2015kxa}, and so the analysis below helps to contrast $\spatinf$ with those cases. We start with an analysis of the free theory and then proceed to interactions.  Then  we study how the bulk Lagrangian can be rewritten in terms of the de Sitter component fields. We also study a specific example of the $\phi^3$ interaction and show how the bulk action induces complicated couplings between these component fields.

\subsection{Asymptotic equations of motion in the free theory}
Before discussing interactions, let us rederive some of our previous results pertaining to a free massive scalar field on Minkowski spacetime that satisfies
\be
\label{classeqfree}
(\square - m^{2})\, \phi(\rho, X)\, =\, 0. 
\ee
We would now like to perform an asymptotic analysis of the space of solutions as we approach $\hat{i}^{0}$ and classify the ``data'' corresponding to the normalizable modes of the massive field at $\spatinf$.
At large $\rho$, the equation of motion reduces to
\be
	\begin{split}
	\left[-{1  \over \rho^2} \partial_{\tau}^2+ \partial_{\rho}^2 +  {3  \over \rho}\partial_{\rho} + { \triangle_{S^2} \over \rho^2 \cosh^2 \tau}  -  {2\tanh \tau \over \rho^2} \partial_\tau - m^2 \right] \phi(\rho,X) = 0.
	\end{split}
\ee
Due to the $m^{2}\phi(\rho, X)$ term, a consistent asymptotic (large $\rho$ at fixed $X$) expansion of the scalar field requires an exponentially decaying tail.\footnote{The most general solution is a linear combination of exponentially growing and decaying modes \cite{Marolf:2006bk}. However here, as in the rest of the paper, we focus on normalizable modes.}  That is,  an ansatz which can be solved recursively in large $\rho$ expansion must have the form
\be
\phi(\rho, X)\, =\, e^{-m\rho}\,\phitail(\rho, X),
\ee
where $\phitail(\rho, X)$ satisfies the following equation.
\be
	\left[-{1  \over \rho^2} \partial_{\tau}^2+ \partial_{\rho}^2 +  \left({3  \over \rho} - 2m \right)\partial_{\rho} - {3 m \over \rho} + { \triangle_{S^2} \over \rho^2 \cosh^2 \tau}  - {2\tanh \tau \over \rho^2} \partial_\tau  \right] \phitail(\rho,X) = 0.
\ee
It can also be checked that the series expansion for $\phitail(\rho,X)$ must start with $\rho^{-{3 \over 2}}$.

Therefore, a natural class of asymptotic falloff conditions that can generate an  infinite-dimensional space of  asymptotic classical solutions to the equations of motion for the massive scalar field is
\be
\phitail(\rho, X)\, =\,  \sqrt{\pi \over 2 m} \frac{1}{\rho^{\frac{3}{2}}}\, \Big[\, \opspat(X)\, +\, \sum_{n=1}^{\infty}\, {1 \over \rho^n} \phitailsub[n](X)\, \Big].
\ee
$\opspat(X)$ is not ``free data'' in the conventional sense since it is not specified on a Cauchy slice but instead on a timelike boundary. Nevertheless, if one thinks of evolution in the $\rho$-coordinate, in the spirit of the Fefferman-Graham expansion in AdS, then the sub-leading term in the $\frac{1}{\rho}$ expansion can be determined in terms of $\opspat(X)$.

This procedure should be applicable in the presence of interactions, which simply add a nontrivial right hand side to the equation above. This can be used to construct subleading terms in $\phitail$ systematically. 

As an example, in the free theory --- where, of course, the equation above can be directly solved --- the coefficient of $\frac{1}{\rho^{\frac{5}{2}}}$ term can be expressed in terms of the leading term as,
\be
\phitailsub[1](X)\, =\, \frac{1}{2m}\, [\, (\partial_{\tau}\, +\, 2\, \tanh\tau\, )\, \partial_{\tau}\, +\, \frac{3}{4}\, -\, \frac{1}{\cosh^{2}\tau}\, \triangle_{S^{2}}\, ]\, \opspat(X) . 
\ee

The classical analysis above recovers the result that $\opspat(X)$ can be defined via ``an extrapolate formula" 
\be
\opspat(X)\, =\, \lim_{\rho\, \rightarrow\, \infty}\, D(\rho)  \phi(\rho, X). 
\ee
The need for smearing, which was discussed at length in section \ref{s_dsfree}, is not evident in the analysis above because we are effectively studying one-point functions. Two-point functions can be easily studied by adding a source to the right hand side of \eqref{classeqfree}, which would make the need for smearing evident again. We do not perform this analysis here since the results so obtained coincide with those of section \ref{s_dsfree}.

We emphasize an  important difference between  the extrapolate dictionary at $\spatinf$ and the extrapolate dictionary at ${\cal I}^{\pm}$ in the massless case.  In the latter case, there is a well-defined notion of the flux of energy-momentum through a cut at null infinity.   As the decay of massive field at $\spatinf$ is exponential we have, 
\be
\lim_{\rho\, \rightarrow\, \infty}\, \int_{S^{2}}\, \rho^{3}\, d^{2} \Omega \, T_{\tau\tau}(\rho, \tau, \Omega)\, =\, 0,
\ee
where $T$ is the stress-tensor of the field. 
The extrapolate dictionary in the massive case requires an exponentially large rescaling factor to determine the boundary fields, which is much larger than the area of the sphere as $\rho \rightarrow \infty$.  

For the same reason, there is no sense in which the bulk symplectic form induces a symplectic form on the data at $\spatinf$. A corollary of this is that a brute force approach to ``asymptotic quantization'' at $\spatinf$ does not work. In fact, due to the potential divergences in extrapolated correlators discussed previously, it is easy to check that the commutator of two extrapolated fields $\opspat(X_1)$ and $\opspat(X_2)$ is not well defined unless both the operators are appropriately smeared.

\subsection{Asymptotic equations of motion for  an interacting scalar field}\label{oEPsimf}

We now turn to the correction to the asymptotic operators at $\spatinf$ induced by the  perturbative self interactions and work out the corrections to the extrapolated operator $\opspat(X)$. The analysis here is performed using the equations of motion, and the final result is presented in momentum space, which is complementary to the perspective of section \ref{s_interactions}.
In order to keep the analysis as general as possible, we study the equations of motion for the massive scalar $\phi(x)$ that interacts locally with an arbitrary set of other fields $\{\Psi\}$. 

Before proceeding with the specific example, we first note that perturbative expansion of a field close to asymptotic boundaries (${\cal I}$ in the case of massless field and $\spatinf$  in the massive case) requires some care. Consider the quadratic perturbation  ${1 \over 2} \lambda^{2}\phi^{2}$. This perturbation can be trivially ``resummed" and the resulting solution is obtained by simply shifting the mass $m^{2}\, \rightarrow\, m^{2} + \lambda^{2}$. Clearly, the new solution, $\phi_{\lambda}$ does not admit any perturbative expansion in $\lambda^{2}$ at large $\rho$ as, 
\be
	\phi_{\lambda}(\rho, X)\, \rightarrow\, \sqrt{\pi \over 2} \frac{1}{(m^{2} + \lambda^{2} )^{1 \over 4} \rho^{\frac{3}{2}}}\, e^{-\sqrt{m^{2} + \lambda^{2}}\, \rho}\, \opspat(X). 
\ee

We now proceed to analyse the effect of higher order local interactions  on the extrapolate limit. We will argue that in the interacting case, the (smeared) asymptotic operator admits a consistent perturbative expansion as, 
\be
	\opspat(X)\, =\, \opspatpert[0](X)\, +\, \lambda\, \opspatpert[1](X)\, +\, O(\lambda^{2}). 
\ee
This expansion is subject to the potential divergences discussed in section \ref{s_interactions} that must be resolved by appropriate smearing.

The interacting equation of motion for  $\phi$ is, 
\be
(\, \square\, -\, m^{2}\, )\, \phi(x)\, =\, \intpot(\phi, \{\Psi\}, x), 
\ee
where $\intpot$ depends on the details of the interaction; for a non-derivative coupling it is simply the derivative of the potential with respect to $\phi(x)$.

The leading order perturbative correction to the free field solution is, 
\be\label{pctffs}
\phipert[1](\rho, X)\, =\, \int\, d^{4}y\, G_{R}(\rho, X, y)\, \intpot(\phipert[0], \{\Psi^{(0)}\}, y), 
\ee
where the superscript $(0)$ indicates that we insert the free-field solutions while evaluating the ``source" $U(\phi,\Psi, y)$. $G_R$ denotes the retarded propagator which is the appropriate choice in the classical theory. $G_R$ is also the appropriate choice of propagator for computing Wightman functions in quantum field theory in leading order in the interaction.  

The retarded propagator in momentum space can be written as
\be
G_{R}(\rho, X,  y)\, =\, \int\, [dl]\, \frac{e^{-i l \cdot (\rho X - y)}}{l^{2} + m^{2}}, 
\ee
where $l^{2}\, :=\, -(l_{0} + i\epsilon)^{2}\, +\, \vec{l}^{2}$ and $[dl]\, =\, -i \frac{d^{4}l}{(2\pi)^{4}}$.
We now use the Kontorovich-Lebedev transform for the plane wave basis, which is
\be
e^{-i (\omega_{l}x^{0} - \vec{l}\cdot\, \vec{x})}\, =\, \frac{1}{\rho}\, \int_{0}^{\infty}\, d\nu\, K_{i\nu}(m\rho)\,  {-2 i \nu \over \pi} C^{(1)}_{-1 + i \nu}(- i \frac{l \cdot X}{m}). 
\ee
Thus $\phipert[1](\rho, X)$ can be written as, 
\be
\begin{split}
\phipert[1](\rho, X)\, &=\, \int_{0}^{\infty}\, d\nu\, \, K_{i\nu}(m\rho)\, \int\, [dl]\, d^{4}y\, {-2 i \nu \over \pi} C^{(1)}_{-1 + i \nu}(-i \frac{l\, \cdot X}{m})\, \frac{e^{i l \cdot y}}{l^{2} + m^{2}}\, \intpot(\phipert[0], \{\Psi^{(0)}\}, y)\\
&=\, \int_{0}^{\infty}\, d\nu\, \, K_{i\nu}(m\rho)\, \int\, [dl]\,  {-2 i \nu \over \pi} C^{(1)}_{-1 + i \nu}(-i \frac{l\, \cdot X}{m})\, \frac{1}{l^{2} + m^{2}}\, \widetilde{\intpot}(\phipert[0], \{\Psi^{(0)}\}, l),
\end{split}
\ee
where $\widetilde{U}$ is the Fourier transform of $U$.
To leading order in perturbation theory, this result is valid for any local interaction and expresses the leading perturbative correction to the free field in terms of the Gegenbauer function. We can decompose $\phipert[1](\rho, X)$ as,
\be
\phipert[1](\rho, X)\, =:\, \frac{1}{\rho}\, \int_{0}^{\infty}\, d\nu\, K_{i\nu}(m\rho)\,\phinupert[1](X), 
\ee
where, 
\be\label{phinu1xcap}
\phinupert[1](X)\, =\, 
\, \int\, [dl]\, \frac{1}{l^{2}+m^{2}}\, {-2 i \nu \over \pi} C^{(1)}_{-1 + i \nu}(-i \frac{l \cdot X}{m})\, \widetilde{\intpot}(l) . 
\ee

The expression for $\phinupert[1](X)$ is an example of a general formula that can be used to go from the operator in momentum space, $\tilde{\phi}(l)$, to the de Sitter component fields, $\phi_{\nu}(X)$.
\be
\phi_{\nu}(X)\, =\, 
\, \int\, [dl]\, {-2 i \nu \over \pi} C^{(1)}_{-1 + i \nu}(-i \frac{l \cdot X}{m})\, \tilde{\phi}(l) . 
\ee
This formula holds for free fields and also for terms that are higher order in $\lambda$.

The correction to the extrapolated operator now takes the form

\be
\label{opspat1correction}
\opspatpert[1](X)\, =\, \lim_{\rho\, \rightarrow\, \infty}\, D(\rho) \,  \phipert[1](\rho, X) = \int_{0}^{\infty} \phinupert[1](X) d \nu. 
\ee
This formula itself does not guarantee that the $\nu$-integral on the right hand side of \eqref{opspat1correction} converges. The Gegenbauer polynomial grows exponentially at large $\nu$. Therefore, as in the free theory, to make sense of this correction it is necessary to either smear the operator or evaluate its matrix elements between appropriate states with bounded $\nu$-content. We leave further study of this issue to future work.

\subsection{Bulk Lagrangian in terms of dS$_{3}$ fields}\label{interdssundrum} 
In this section, we decompose the Lagrangian in terms of component de Sitter fields. Our objective is to show that, at the free-level, the Lagrangian breaks up into a sum over Lagrangians for each component but when interactions are added, they cause different de Sitter fields to mix with each other.
The analysis here is similar to the one in \cite{Sundrum:2011ic}, where a free field in $D$ dimensional AdS was decomposed in terms of a one parameter family of free fields on $D-1$ dimensional Minkowski space. 

We find the Lagrangian more convenient to study than the action. This is because the  Lagrangian is given by the integral of the Lagrangian density over a Cauchy slice,  which can be entirely described in the coordinate patch that we have studied in this paper.

\subsubsection{Free field}
The Lagrangian density for a free field is given by
\be
{\cal L} = -{1 \over 2} \partial_{\mu} \phi \partial_{\nu} \phi g^{\mu \nu} - {1 \over 2} m^2 \phi^2 = -{1 \over 2} \partial_{\rho} \phi \partial_{\rho} \phi -{1 \over 2} \partial_{a} \phi \partial_{b }\phi \, g^{ab} - {1 \over 2} m^2 \phi^2, 
\ee
where $\mu, \nu$ run over all indices and $a,b$ run over the de Sitter directions. We study its integral over a Cauchy slice, which yields the Lagrangian
\be
\label{ldef}
L = \int {\sqrt{g}} {\cal L} d \rho d \Omega = -\cosh^2 \tau \int \rho^3 d \rho d \Omega \left[{1 \over 2} \partial_{\rho} \phi \partial_{\rho} \phi +{1 \over 2} \partial_{a} \phi \partial_{b }\phi \, g^{ab} + {1 \over 2} m^2 \phi^2 \right]. 
\ee

We now insert the decomposition \eqref{dsdecomp} into the Lagrangian. Note that the decomposition \eqref{dsdecomp} can be performed entirely off-shell and does not require that $\phi$ obeys the equation of motion.  
Integrating by parts in the term involving the $\rho$-derivative, using the defining equation for the Macdonald function and the orthogonality relation \cite{passian2009orthogonality} that holds when $\nu, \nu' > 0$  
\be
\label{besselorthogonality}
\int_0^{\infty} K_{i \nu} (x) K_{i \nu'}(x)  {d x \over x} = \delta(\nu - \nu') {\pi^2 \over 2} {1 \over \nu \sinh \pi \nu}, 
\ee
we find that
\be
\label{ldecompose}
L =  \int_0^{\infty} d \nu {\pi^2 \over 2 \nu \sinh \pi \nu} L_{\nu}, 
\ee
where $L_{\nu}$ is the Lagrangian for a field with mass parameter $\nu$ in de Sitter space
\be
L_{\nu} = -\int \cosh^2 \tau d \Omega \left[{1 \over 2} g^{a b}_{\dsthree} \partial_{a} \phi_{\nu} \partial_{b} \phi_{\nu}  + {1 \over 2} (1 + \nu^2) \phi_{\nu}^2 \right]. 
\ee
Here $g_{\dsthree}^{a b}$ is the inverse of the metric on de Sitter space \eqref{blowupmetric} which differs from the inverse of the Minkowski metric since it has no factor of ${1 \over \rho^2}$.

The conclusion above is that the Lagrangian for the field decomposes into a linear sum of the Lagrangians for its constituent fields. Note that the leading $\nu$-dependent factor in \eqref{ldecompose} can be absorbed into the normalization of the $\phi_{\nu}$ fields. Indeed, this normalisation is consistent with the one obtained from the  Klein-Gordon inner product in \eqref{eucmodlnorm}.
 
\subsubsection{Interacting field \label{subsecinteractlag}}
However, in the presence of interactions,  it is {\em not} the case that the Lagrangian is the sum of interacting Lagrangians for individual component fields in de Sitter space. Instead, interactions typically couple fields with different masses in de Sitter. This can be seen explicitly in the example of a simple  cubic self-interaction $\lambda\, \phi(\rho, X)^{3}$.  

The interaction Lagrangian  can be written as,
\be
L_{\text{int}}  =\, {\lambda} \cosh^2 \tau \int\,   d \Omega \, \int \, f(\nu_{1}, \nu_{2}, \nu_{3})\, \phi_{\nu_{1}}(X)\, \phi_{\nu_{2}}(X)\, \phi_{\nu_{3}}(X) d \nu_1 d \nu_2 d \nu_3, 
\ee
where, 
\be
\label{fnuidef}
f(\nu_{1}, \nu_{2}, \nu_{3})\, =\,  \int_{0}^{\infty}\, d\rho\, \prod_{j=1}^{3}\, K_{i\nu_{j}}(m\rho). 
\ee
The ``recoupling'' scheme for three Macdonald functions is known. An application of the formula in section 2.16.46 in \cite{Prudnikov_book-II} yields
\be
\label{fnuiform}
f(\nu_i)\, =\,{1  \over m} \, \sum_{\sigma_{1}, \sigma_{2}\,= \pm 1}\, {1 \over 8} A(\sigma_{1}\nu_{1}, \sigma_{2}\nu_{2}),
\ee
where 
\be
\begin{split}
A(\nu_{1}, \nu_{2})\, &= \Gamma[\, -i\nu_{1}] \Gamma[-i\nu_{2}] \Gamma[\frac{1 + i(\nu_{1} + \nu_{2} - \nu_{3})}{2}] \Gamma[\frac{1 + i(\nu_{1} + \nu_{2} + \nu_{3})}{2}\, ]\, \\ &\times F_{4}( \frac{1 + i(\nu_{1} + \nu_{2} - \nu_{3})}{2},\, \frac{1 + i(\nu_{1} + \nu_{2} + \nu_{3})}{2}\,; 1 + i\nu_{1}, 1 + i\nu_{2};\, 1,\, 1),
 \end{split}
\ee
and $F_{4}$ is one of the four Appell hypergeometric functions $F_{4}(a, b, c,d;, x, y)$. The Appell hypergeometric series is not, in general, convergent at $x = y = 1$, but the case at hand simplifies since we have $a + b + 1= c+d$. We can then use a suitable analytic continuation of the formula given in \cite{bailey1933reducible} to write it in terms of a product of two $_2F_1$ hypergeometric functions. This yields
\be
\label{finalsimpA}
\begin{split}
A(\nu_{1}, \nu_{2})\, &= \Gamma[\, -i\nu_{1}] \Gamma[-i\nu_{2}] \Gamma[\frac{1 + i(\nu_{1} + \nu_{2} - \nu_{3})}{2}] \Gamma[\frac{1 + i(\nu_{1} + \nu_{2} + \nu_{3})}{2}\, ]\, \\ &\times \, _2F_{1}( \frac{1 + i(\nu_{1} + \nu_{2} - \nu_{3})}{2},\, \frac{1 + i(\nu_{1} + \nu_{2} + \nu_{3})}{2}\,; 1 + i\nu_{1};e^{i {\pi \over 3}})  \\ &\times \, _2F_{1}( \frac{1 + i(\nu_{1} + \nu_{2} - \nu_{3})}{2},\, \frac{1 + i(\nu_{1} + \nu_{2} + \nu_{3})}{2}\,; 1 + i\nu_{2};e^{i {\pi \over 3}}).
\end{split}
\ee
The equality between \eqref{fnuidef} and \eqref{fnuiform} can be verified numerically using the expression \eqref{finalsimpA}.

This analysis shows how integrating over the $\rho$ coordinates results in rather complicated $\textrm{dS}_{3}$ interactions that mix modes with different values of $\nu$.

\section{Discussion \label{s_conclusions}}
In this paper, we have explored a novel extrapolation procedure for massive quantum fields in four-dimensional flat space that leads to well-defined asymptotic observables. These observables live on an asymptotic  $\dsthree$ slice, which we denote by  $\spatinf$. Our extrapolation procedure relies on the observation that massive fields have exponential tails as one approaches $\spatinf$. In the interacting theory, the leading falloff is controlled by the renormalized mass of the field. Stripping off these tails yields operators intrinsic to $\spatinf$.

The boundary operators obtained through this procedure are subtle, and their correlators might display infrared divergences. We argued that, in the free theory, these divergences can be  cured by smearing the operators appropriately.  

We obtained a number of interesting results related to the algebra of boundary operators.  We showed that local bulk fields can be obtained by smearing the boundary operators over $\spatinf$ with carefully chosen smearing functions in the free theory. This is analogous to the HKLL reconstruction in AdS/CFT. We also derived a general result, valid in both the interacting and the free theory, that relates the width of the smearing function to the localization of the bulk field. Although our choice of an origin breaks translational invariance, we showed that boundary fields transform simply under a set of global spacetime translations.

We took some initial steps towards understanding the impact of interactions on boundary correlators. This remains an important direction for future work. In particular, we would like to understand whether our procedure for smearing boundary operators is sufficient to regulate potential divergences in interacting correlators, or whether a more refined procedure is required.

In forthcoming work  \cite{Laddha:2022masshol}, we will present support for the claim that  $\spatinf$ is relevant for flat-space holography in theories with dynamical gravity. We will show that, in gravitational theories,  $\spatinf$ stores information holographically. In particular, observations from a restricted subalgebra on $\spatinf$ when combined with observations near  $\scrippast$ are sufficient to uniquely identify any state of the theory.

An interesting aspect of this description, which is clear from our analysis of interactions, is that the dynamics at $\spatinf$ is nontrivial. This is in contrast to the description at $\scrip$ and $i^{+}$ (or $\scrim$ and $i^{-}$) where (barring separate subtleties having to do with IR-divergences in four dimensions) the dynamics becomes free. But it is similar to the boundary of AdS, where boundary dynamics is nontrivial. This is a welcome feature and we expect that it will aid us in understanding flat-space holography since the nontriviality of AdS boundary dynamics is what encodes the bulk dynamics in AdS/CFT.

It would also be interesting to understand the relation of correlators on $\spatinf$ to correlators on the celestial sphere that have been studied extensively in the recent literature \cite{Fan:2020xjj ,Albayrak:2020saa,Law:2020xcf,Raclariu:2021zjz,McLoughlin:2022ljp,Pasterski:2021raf,Donnay:2022aba,Donnay:2021wrk}. 

We would like to conclude by pointing out that the questions we have explored in this paper are very new. We have only scratched the surface of what, we believe, is a very interesting structure that will be useful both in studies of quantum field theory and of quantum gravity in asymptotically Minkowski space.

\section*{Acknowledgments}
We are grateful to Anupam A H, Miguel Campiglia, Bidisha Chakrabarty,  Tuneer Chakraborty, Joydeep Chakravarty, Chandramouli Chowdhury,  Abhijit Gadde, Hao Geng, Victor Godet, Diksha Jain, Andreas Karch, Suman Kundu, Gautam Mandal, Shiraz Minwalla, Kyriakos Papadodimas, Olga Papadoulaki, Priyadarshi Paul, Carlos Perez-Pardavila, Onkar Parrikar, Athira P V, Lisa Randall, Marcos Riojas, Sanjit Shashi, Sandip Trivedi, Amitabh Virmani and Edward Witten  for helpful discussions.  We would like to thank Semyon Yakubovich for patiently answering our queries pertaining to index transforms.  S.R. is partially supported by a Swarnajayanti fellowship, DST/SJF/PSA-02/2016-17, of the Department of Science and Technology. Research at ICTS-TIFR, and at TIFR is supported by the Department of Atomic Energy, Government of India, under Project Identification Nos. RTI4001 and RTI4002 respectively.  S.P. acknowledges support from a J C Bose Fellowship JCB/2019/000052. S.P. and P.S acknowledge support from the Infosys Endowment for the study of the Quantum Structure of Spacetime. S.R. is grateful to the Frontier Symposium in Physics (IISER Thiruvananthapuram) and the 12\textsuperscript{th} Crete Regional Meeting in String Theory for hospitality while this work was being completed. P.S. acknowledges the support of the Centre for High Energy Physics at the Indian Institute of Science, where part of this work was completed.

\appendix

\section{More details on the dS decomposition \label{app_moredetails}}

In this appendix we provide details for some of the results pertaining to the de Sitter decomposition of the free massive scalar field described in section \ref{s_dsfree}.

\subsection{Details of the conformal basis}\label{app_confbasis}
Recall that the free massive scalar field can be expanded in the conformal basis \eqref{modeexpconfp}. The goal of this part of the appendix is to determine the correct normalization and derive an orthogonality relation satisfied by the conformal  mode functions. In what follows, we allow $d$ to be arbitrary, even though we are interested in $d=3$ in the main text.
\subsubsection*{Normalization}
In order to determine the normalization such that the oscillators obey canonically commutation relations \eqref{bbdaggercomm}, it is sufficient to compute the vacuum two-point function in this basis and match it with the known result in Minkowski space,
\be
\langle 0 | \phi(\rho,X) \phi(\rho',X') |0\rangle = { m^{d-1\over 2} \over (2\pi)^{d+1\over 2} } {K_{d-1\over 2}(m \lambda) \over \lambda^{d-1\over 2}}, \quad \lambda^2 = (\rho X - \rho' X')^2,
\ee
where $\lambda$ is the geodesic distance between the two insertions.
On the other hand, in the conformal basis, using the expansion \eqref{modeexpconfp} and commutation relations \eqref{bbdaggercomm}, we get 
\be
\begin{split}
	\langle 0 | \phi(\rho,X) \phi(\rho',X') |0\rangle & = \int\limits_0^{\infty} d\nu \dmuxi |N_{\nu}|^2 {K_{i\nu}(m \rho) \over \rho^{d-1\over2} } \, {K_{i\nu}(m \rho') \over (\rho')^{d-1\over2} } \, \psi_{\nu,\xi}(X) \bar{\psi}_{\nu,\xi}(X').
\end{split}
\ee
These two results can be seen to be equal through the following identity \cite{durand1979addition} (see equation 19a),
\be
{K_{d-1\over 2}(w) \over w^{d-1\over 2}} = - {2^{d-1\over 2} \over \pi} \Gamma({d-1\over2}) \int\limits_{0}^{\infty} {K_{i\nu}(x)\over x^{d-1\over 2}}{K_{i\nu}(y)\over y^{d-1\over 2}} {C^{d-1\over 2}_{i\nu - {d-1\over 2}}(-z) \over \sin(\pi(i\nu - {d-1\over 2}))} \sinh(\pi \nu) \nu d\nu,
\ee
where
\be
w^2 = x^2 + y^2 -2 x y z  . 
\ee
The equivalence of the two point function computed from plane wave mode expansion and conformal mode expansion requires
\be\label{normfromtpf}
|N_{\nu}|^2 \int \dmuxi  \, \psi_{\nu,\xi}(X) \bar{\psi}_{\nu,\xi}(X') = -{\Gamma({d-1\over2})\over 2 \pi^{d+3\over 2} } {\nu \sinh(\pi \nu) C^{d-1\over 2}_{i\nu - {d-1\over 2}}(-X \cdot X') \over \sin(\pi(i\nu - {d-1\over 2}))} .
\ee
The left hand side can be computed by extending the coordinates $X$ by adding a past-directed timelike imaginary part and $X'$ by adding a future-directed timelike imaginary part, as discussed earlier. This is the two-point Wightman function of a free massive scalar field on de Sitter space \cite{Bros:1995js}. In embedding coordinates, it is a function of $X\cdot X'$ and can be computed by making the following choices.
\be
\begin{split}
&X = i(-1,0,0\cdots,0) , \  X' = i(\cosh v ,-\sinh v,0\cdots,0), \  \xi = (1,\cos \theta, \sin\theta\cos\phi,\cdots),\\
&X \cdot X' = -\cosh v , \quad -X \cdot \xi = e^{-i {\pi \over 2}}, \quad -X'\cdot \xi =  e^{i {\pi \over 2}} (\cosh v+ \sinh v\cos\theta).
\end{split}
\ee
With these choices, the integral simplifies to
\be
\begin{split}
&  e^{-{\pi\over 2} (\nu+\nu')}  \omega_{d-1} \int\limits_{0}^{\pi} d\theta (\sin \theta)^{d-2} (\cosh v+ \sinh v\cos\theta)^{-{d-1\over 2} + i \nu'} \\
&=  2\pi^{d/2}{\Gamma(d-1)\over \Gamma({d\over 2})}  {\Gamma(-{d-3\over2}+i\nu') \over \Gamma({d-1\over2}+i\nu')}  e^{-{\pi\over 2} (\nu+\nu')} C^{d-1\over2}_{-{d-1\over2}+i\nu'}(-X\cdot X').
\end{split}
\ee
We have denoted the surface area of $S^{d-1}$ by $\omega_d$ and used the integral representation of the Gegenbauer function \cite{Batemanbook}.
We fix the normalization by substituting the integral above in \eqref{normfromtpf}.
\be
|N_{\nu,\vec{w}}|^2 = { \nu \Gamma({d-1\over2}+i\nu) (e^{2\pi\nu}-1)\over (2\pi)^{d+1} \Gamma(-{d-3\over2}+i\nu) \cos(\pi({d\over2}-i\nu))}.
\ee
When $d=3$, i.e., when the Minkowski spacetime is four dimensional, the normalization takes the simple form,
\be
N_{\nu,\vec{w}} = {\nu \, e^{\pi\nu\over 2}  \over  2^{3/2} \, \pi^2}.
\ee

\subsubsection*{Orthogonality relations}
The conformal modes satisfy an orthogonality relation, which is most conveniently expressed in the following coordinate system. If we parameterize the points on de Sitter as
\be\label{embedspacelike}
X = \left({1-y^2 + |\vec{z}|^2 \over 2 y}, {\vec{z}\over y} , {1+y^2 - |\vec{z}|^2 \over 2 y} \right) , 
\ee
and the null vectors as
\be\label{embedtimelike}
q(\vec{w} ) = (1+|\vec{w}|^2 , 2\vec{w}, 1- |\vec{w}|^2),
\ee
then the orthogonality relation can be written as
\be\label{dsorthogonality}
\begin{split}
&\int \dds \, (-X\cdot q(\vec{w}) - i\epsilon)^{-{d-1\over2}-i\nu} \, (-X\cdot q(\vec{w\,}') - i\epsilon)^{-{d-1\over2}-i\nu'}  \\
&  = {2\pi^d \Gamma(i\nu)\Gamma(-i\nu)\over \Gamma({d-1\over 2}+i\nu)\Gamma({d-1\over 2}-i\nu)} \delta(\nu+\nu')\delta(\vec{w}-\vec{w\,}') + \frac{ 2\pi^{\frac{d+1}{2}} \Gamma(i\nu)}{\Gamma({d-1\over 2}+i\nu)}  e^{-\pi \nu}  {\delta(\nu-\nu')\over |\vec{w}-\vec{w\,}'|^{2+2i\nu}},
\end{split}
\ee
where $\dds = dy d^{d-1} \vec{w} / y^{d}$ denotes the invariant measure on the de Sitter space. To compute this integral, it is convenient to introduce the following Fourier transform.\footnote{This computation is motivated by an analogous computation in AdS \cite{Costa:2014kfa}.}
\be
\begin{split}
& \Pi_{\nu}(y,\vec{k})  \defeq \int d^{d-1} \vec{z} \, {e^{i\vec{k}.\vec{z}} \over \left( |\vec{z}|^2 - y^2 - i\epsilon \right)^{{d-1\over 2}+i\nu}}; \\
&= \int_0^{\infty} {dt \over t} \, {t^{{d-1\over 2}+ i \nu} \over \Gamma({d-1\over 2}+i\nu)} \int d^{d-1}\vec{z} \, e^{i\vec{k}\cdot\vec{z}}\, e^{- (|\vec{z}|^2 - y^2 -i \epsilon) t} = {\pi^{d-1 \over 2} \over \Gamma({d-1\over 2}+i\nu)}  \int_0^{\infty} {dt \over t} \, t^{ i \nu} e^{-\left( (- y^2 -i\epsilon) t + {k^2\over 4t}\right)}.
\end{split}
\ee
Substituting in the LHS of \eqref{dsorthogonality} we get,
\be
\begin{split}
& \int_0^{\infty} {dy\over y} d^{d-1}\vec{z} y^{i\nu + i\nu'} \int {d^{d-1} \vec{k} d^{d-1}\vec{k\,}'\over (2\pi)^{2d-2}} e^{-i\vec{k}.(\vec{z}-\vec{w})}\,e^{-i\vec{k}'.(\vec{z}-\vec{w}')} \\
& \times  {\pi^{d-1} \over \Gamma({d-1\over 2}+i\nu) \Gamma({d-1 \over 2}+i\nu')} \int_0^{\infty} {dt \over t} \, t^{ i \nu} e^{-\left( (- y^2 -i\epsilon) t + {k^2\over 4t}\right)}  \int_0^{\infty} {dt' \over t'} \, (t')^{ i \nu'} e^{-\left( (- y^2 -i\epsilon) t' + {k'^2\over 4t'}\right)}\\
&= \int_0^{\infty} {dy\over y}y^{i\nu + i\nu'} \int {d^{d-1} \vec{k}\over (2\pi)^{d-1}} e^{i\vec{k}.(\vec{w}-\vec{w\,}')} {\pi^{d-1} \over \Gamma({d-1\over 2}+i\nu) \Gamma({d-1\over2 }+i\nu')}\\
&\times  \int_0^{\infty} {dt \over t} \, t^{ i \nu} e^{-\left( (- y^2 -i\epsilon) t + {k^2\over 4t}\right)}  \int_0^{\infty} {dt' \over t'} \, (t')^{ i \nu'} e^{-\left( (- y^2 -i\epsilon) t' + {k'^2\over 4t'}\right)}.
\end{split}
\ee
The $t$-integrals give us Macdonald functions using the relation
\be
\int {dt\over t} t^{i\nu} e^{-\eta^2 t - {k^2\over 4t}} = 2 \left({k\over 2 \eta}\right)^{i\nu} K_{i\nu}(k \eta),
\ee
with
\be
\eta^2 = -y^2 - i\epsilon \equiv  |y|^2 e^{-i \pi} \quad\implies \eta = e^{-i{\pi\over 2}} |y|. 
\ee

Hence,
\be
\begin{split}
&\int_0^{\infty} {dy\over y^d} d^{d-1}\vec{z} {y^{{d-1 \over 2}+ i \nu} \over (|\vec{z}-\vec{w}|^2 - y^2 - i\epsilon)^{{d-1 \over 2}+i\nu}}{y^{{d-1 \over 2}+ i \nu'} \over (|\vec{z}-\vec{w\,}'|^2 - y^2-i\epsilon)^{{d-1\over 2}+i\nu'}} \\
&=\int_0^{\infty} {dy\over y} y^{i\nu + i\nu'}  \int {d^{d-1} \vec{k}\over (2\pi)^{d-1}} e^{i\vec{k}.(\vec{w}-\vec{w\,}')} {\pi^{d-1} \over \Gamma({d-1\over 2}+i\nu) \Gamma({d-1\over2 }+i\nu')} \,  \\& \hspace{15 pt} \times 4\left({k\over 2 e^{-i{\pi\over 2}} |y|}\right)^{i\nu}  \left({k\over 2 e^{-i{\pi\over 2}} |y|}\right)^{i\nu'} K_{i\nu}(- i k |y| )  K_{i\nu'}(- i k |y| ) \\
&=\int {d^{d-1} \vec{k}\over (2\pi)^{d-1}}  {e^{i\vec{k}.(\vec{w}-\vec{w\,}')} \, \pi^{d-1} \over \Gamma({d-1\over 2}+i\nu) \Gamma({d-1\over2 }+i\nu')}\,  4 e^{-\pi{\nu+\nu'\over 2}} \left({k\over 2 }\right)^{i(\nu+\nu')}  \int_0^{\infty} {dy\over y}   K_{i\nu}(- i y )  K_{i\nu'}(- i y ),
\end{split}
\ee
where in the last line we rescaled $|y|\rightarrow |y| /k$. The integral over $y$ can be regulated and gets a contribution only when $\nu=\pm \, \nu'$.
\be
\begin{split}
\lim\limits_{\epsilon\rightarrow 0 } \int_0^{\infty} {dy\over y} y^{\epsilon}  K_{i\nu}(- i y )  K_{i\nu'}(- i y ) = {\pi \Gamma(i\nu)\Gamma(-i\nu) \over 2} \left( \delta(\nu+\nu')+ \delta(\nu-\nu')\right).
\end{split}
\ee
Note that we have an additional term compared to \eqref{besselorthogonality} since here $\nu, \nu'$ can be both positive and negative. Finally, we can perform the $k$ integral to obtain the RHS of \eqref{dsorthogonality}, thereby completing the proof.

\subsection{Spherical basis on $\spatinf$ \label{appsphbasis}}

In this section, we will describe the spherical basis used to expand the de Sitter free field $\phi_{\nu}(X)$ and also work out the correct normalization of $\phi_{\nu}(X)$ when it appears as a component of the bulk field. We recall, from \eqref{dseqn} in the main text, that it satisfies the equation
\be
(\Box_{\dsthree} - (1 + \nu^2)) \phi_{\nu}(X) = 0.
\ee
Working with the metric in the global coordinates,
\be
ds_3^2 = -d \tau^2 + \cosh^2 \tau d \Omega^2,
\ee
we decompose the massive field into spherical harmonics on the $S_2$. We can write the mode decomposition for the classical field as
\be
\phi_{\nu}(X) = \sum_{\vec{\ell}} {\tN_{\nu }}\, \func_{\nu, \ell}(\tau) Y_{\vec{\ell}}\,(\Omega) + \hc,
\ee
with $\func_{\nu,\ell}(\tau)$ being the solution to
\be
\label{tauequationsph}
\left(\partial^2_\tau +2 \tanh\tau \, \partial_\tau +\ell(\ell+1)\text{sech}^2 \tau \, +1+\nu^2\right) \func_{\nu,\ell}(\tau)=0. 
\ee
We need to find the correct independent solutions to this equation. 

A change of variables from $\tau \rightarrow \cosh^2 \tau$ converts \eqref{tauequationsph} into a hypergeometric equation, to which a set of independent solutions are 
\be
\begin{split}
F_{1;\nu,\ell}(\tau)&= (\cosh\tau)^{-i \nu-\ell}\,  _2F_1 \left(\half (i \nu-\ell),\half (1+i \nu+\ell);1+i \nu;\frac{1}{\cosh^2\tau}\right);\\
F_{2;\nu,\ell}(\tau)&= (\cosh\tau)^{i \nu-\ell}\,  _2F_1 \left(-\half (i \nu+\ell),\half (1-i \nu+\ell);1-i \nu;\frac{1}{\cosh^2\tau}\right),
\end{split}
\ee
with  $\nu \in \mathbb{R}$. Note that $F_{2;\nu,\ell}(\tau)=F_{1;\nu,\ell}^*(\tau)=F_{1;-\nu,\ell}(\tau)$ and, in what follows, we restrict attention to $\nu > 0$.   The set of solutions above do not form a complete basis for the entire range of $\tau$ since, in the form above, they are symmetric under $\tau \rightarrow -\tau$. 

By a set of hypergeometric transformations, we can transform the solutions above into a form that is manifestly complete.  We start with a somewhat obscure hypergeometric transformation (see \cite[Eq.~15.8.20]{NIST:DLMF}; \cite[Eq.~15.8.15]{NIST:DLMF} gives an alternate expression)
\be
_2F_1({c - a \over 2}, {a + c - 1 \over 2}, c, 4 z (1 - z)) = (1- z)^{1 - c} \, _2F_1(a, 1-a,c, z), \qquad \text{Re}(z) < 1.
\ee
We use this identity with the following values
\be
c = 1 + i \nu; \quad a = \ell + 1; \quad z = {1 \over 2}(1 - \tanh \tau), 
\ee
where we have assumed that $\tau > 0$. With these values, we also have 
\be
(1 - z)^{1 - c} = (\cosh \tau + \sinh \tau)^{-i \nu} (\cosh \tau)^{i \nu}. 
\ee
The two solutions above can be transformed into 
\be \label{eq:legp}
\begin{split}
p_{\nu,\ell}(\tau)=& \frac{e^{-i \nu  \tau}}{\cosh\tau}\, \,  _2F_1 \left(\ell+1,-\ell;1+i \nu;\half (1-\tanh \tau)\right);\\
\tilde{p}_{\nu,\ell}(\tau)=&  \frac{e^{i \nu \, \tau}}{\cosh\tau}\, \,  _2F_1 \left(\ell+1,-\ell;1-i \nu;\half (1-\tanh \tau)\right).
\end{split}
\ee
These solutions are related to the ones above by
\be \label{eq:ftog}
\begin{split}
F_{1;\nu,\ell}(\tau)&=2^{i \nu}p_{\nu,\ell}(\tau), \qquad \tau > 0;\\
F_{2;\nu,\ell}(\tau)&=2^{-i \nu}\tilde{p}_{\nu,\ell}(\tau), \qquad \tau > 0,
\end{split}
\ee
but the functions do not coincide for negative $\tau$.

As with the first pair, we have $\tilde{p}_{\nu,\ell}(\tau)=p_{\nu,\ell}^*(\tau)=p_{-\nu,\ell}(\tau)$. Therefore we simply work with the solutions $p_{\nu, \ell}(\tau)$ and $p_{-\nu, \ell}(\tau)$ below.

Next, we use the Euler transformation,
\be
_2F_1\left(a,b,c,z\right)=(1-z)^{-b}\,_2F_1(c-a,b,c;\frac{z}{z-1}),
\ee
to work with the following set of hypergeometric solutions to the equation \eqref{tauequationsph}.
\be
\begin{split}
p_{\nu,\ell}(\tau)=& \frac{e^{(\ell-i \nu) \tau}}{2^\ell(\cosh\tau)^{\ell+1}}\, \,  _2F_1 \left(i \nu-\ell,-\ell;1+i \nu;-e^{-2\tau}\right)\\
p_{-\nu,\ell}(\tau)=& \frac{e^{(\ell+i \nu) \tau}}{2^\ell(\cosh\tau)^{\ell+1}}\, \,  _2F_1 \left( -i \nu-\ell,-\ell;1-i \nu;-e^{-2\tau}\right).
\end{split}
\ee 
Although we have made a number of transformations, these are still related to the first set of hypergeometrics via  \eqref{eq:legp} for positive $\tau$.

The hypergeometrics in this pair are actually just finite polynomials as the second argument is a negative integer. This can be made explicit by noting that the hypergeometrics in the form given in \eqref{eq:legp} are associated Legendre polynomials. Using the identity
\be
P^{m}_{n}(z)=\frac{1}{\Gamma(1-n)}\left(\frac{1+z}{1-z}\right)^{m \over 2} \, _2F_1(-m,m+1,1-n;\frac{1-z}{2}),
\ee
we find that the set of independent solutions can be expressed as
\be \label{ptolegp}
\begin{split}
p_{\nu,\ell}(\tau)&=\Gamma(1+i \nu)  \frac{P_{\ell}^{-i \nu}(\tanh\tau)}{\cosh\tau}; \\
p_{-\nu,\ell}(\tau)&= \Gamma(1-i \nu)  \frac{P_{\ell}^{i \nu}(\tanh\tau)}{\cosh\tau}.
\end{split}
\ee

\subsubsection{Euclidean modes } \label{sec:eucmodes}
The correct combination of $p_{\nu, \ell}(\tau)$ and $p_{-\nu, \ell}(\tau)$ that serves as the ``in'' modes is fixed by the Euclidean regularity condition. This means that we analytically continue $\dsthree$ to $S^3$ by taking $\tau \rightarrow i \tau$ and then consider modes that are regular near the south pole of the $S^3$. More precisely, the right linear combination is fixed by the condition \cite{Mottola:1984ar}
\be
\lim_{\tau \rightarrow -{i \pi \over 2}} (\tau + {i \pi \over 2}) \left(p_{\nu, \ell}(\tau) + c_{\nu, \ell} p_{-\nu, \ell} \right) = 0.
\ee
This sets
\be
c_{\nu, \ell} = -e^{-\pi  \nu }\frac{ \Gamma (1+i \nu ) \Gamma (\ell-i \nu +1)}{\Gamma (1-i \nu ) \Gamma (\ell+i \nu +1)}.
\ee

The negative energy modes, which multiply the creation operators are, of course, regular at $\tau = {i \pi \over 2}$ 
\be
\lim_{\tau \rightarrow {i \pi \over 2}} (\tau - {i \pi \over 2}) \left(p^*_{\nu, \ell}(\tau) + c^*_{\nu, \ell} p^*_{-\nu, \ell} \right) = 0.
\ee

In section  \ref{conftospheranalyticity}, we will  show that when the positive/negative modes in the conformal basis are transformed to the spherical basis, they correspond precisely to the positive/negative energy modes defined here by demanding Euclidean regularity:
\be
\begin{split}
\func_{\nu,\ell}(\tau)&= p_{\nu, \ell}(\tau) + c_{\nu, \ell}\, p_{-\nu, \ell}(\tau); \\
\func^*_{\nu,\ell}(\tau)&= p_{-\nu, \ell}(\tau) + c_{\nu, \ell}^*\, p_{\nu, \ell}(\tau).
\end{split}
\ee

\subsubsection{Normalization} 
Let us write the decomposition of the bulk massive scalar field in the spherical basis as 
\be
	\phi(\rho,X) = \int d\nu\, \sum_{\vec{\ell}} \tB_{\nu, \vec{\ell \, }} \Phi^+_{\nu,\vec{\ell}} + \tB^{\dagger}_{\nu, \vec{\ell \, }} \Phi^-_{\nu,\vec{\ell}} 
	=\int d\nu\, \sum_{\vec{\ell}} \tB_{\nu, \vec{\ell \, }} \, \tN_{\nu} \frac{ K_{i \nu}(m \rho)}{\rho}  \func_{\nu,\ell}(\tau) Y_{\vec{\ell}}\,(\Omega) + \hc,
\ee
From the previous discussion of analyticity properties of $\func$, it is clear that $\Phi^+$ are the positive energy modes in Minkowski space. We wish to find the right normalization factor $\tN_{\nu}$ which gives the canonical Klein Gordon norm for $\Phi^{+}_{\nu,\vec{\ell}}$. The latter is given by
\be
	\left( \Phi^+_{\nu,\vec{\ell\,}} \,  , \, \Phi^+_{\nu',\vec{\ell'}} \right)_{\text{KG}} =-i \int_{S}dS \, n^{\mu} (\Phi^+_{\nu,\vec{\ell}}\,\, \partial_\mu \Phi^{+*}_{\nu',\vec{\ell'}}-\Phi^{+*}_{\nu',\vec{\ell'}}\, \partial_\mu \Phi^+_{\nu,\vec{\ell}})=\delta(\nu-\nu')\delta_{\vec{\ell\,},\vec{\ell'\,}},
\ee
where $S$ is some Cauchy surface with normal $n^{\mu}$ and volume element $dS$.
Choosing $S$ to be a constant $\tau$ surface with $\tau=\tau_0$, we get
\be
	\begin{split}
		\left( \Phi^+_{\nu,\vec{\ell\,}} \,  , \, \Phi^+_{\nu',\vec{\ell'}} \right)_{\text{KG}}   =&-i (\func_{\nu,\ell}\partial_\tau \func_{\nu',\ell'}^{*}-\func^{*}_{\nu',\ell'} \partial_\tau \func_{\nu,\ell} )\cosh^2(\tau)|_{\tau=\tau_0} \\
		& \tN_{\nu}\tN^*_{\nu'}\int_{\rho\geq 0} \rho \, d\rho \frac{K_{i\nu}(m\rho)}{\rho} \frac{K_{i\nu'}(m\rho)}{\rho}\int_{S^2} d^2\Omega \, Y_{\vec{\ell},m}(\Omega) Y^{*}_{\vec{\ell}',m'}(\Omega).
	\end{split}
\ee
We now use the orthogonality relation for the Macdonald functions \eqref{besselorthogonality} to do the $\rho$ integral. 
We normalize the spherical harmonics such that
\be
\label{normsphericalharm}
\int d^2\Omega \, Y_{\vec{\ell}}\,(\Omega) Y^{*}_{\vec{\ell'}}(\Omega)=\delta_{\vec{\ell\,},\vec{\ell'\,}}
\ee
so that we get
\be
\begin{split}
	\left( \Phi^+_{\nu,\vec{\ell\,}} \,  , \, \Phi^+_{\nu',\vec{\ell'}} \right)_{\text{KG}} =& -i(\func_{\nu,\ell}\,\partial_\tau \func_{\nu,\ell}^{*}-\func^{*}_{\nu,\ell}\, \partial_\tau \func_{\nu,\ell} )\cosh^2(\tau)|_{\tau=\tau_0} \\
	&\times \frac{\pi^2 |\tN_{\nu}|^2}{2 \nu  \sinh(\pi \nu)} \delta(\nu-\nu')\delta_{\vec{\ell\,},\vec{\ell'\,}}.
	\end{split}
\ee
We find that 
\be\label{eucmodlnorm}
		(\func_{\nu,\ell}\partial_\tau \func_{\nu,\ell}^{*}-\func^{*}_{\nu,\ell} \partial_\tau \func_{\nu,\ell} )\cosh^2(\tau)|_{\tau=\tau_0}=2i \nu   (1-{|c_{\nu,\ell}|^2}) =2i \nu  (1-e^{-2\pi \nu}) .
\ee 
Thus, the normalization factor for these modes is given by
\be\label{eucnorm}
\tN_{\nu} = {e^{\pi \nu \over 2} \over \sqrt{2} \, \pi} .
\ee
Finally, note that the inner product between a positive and a negative energy mode is given by 
\be
		\left( \Phi^+_{\nu,\vec{\ell\,}} \,  , \, \Phi^-_{\nu',\vec{\ell'}} \right)_{\text{KG}} \propto -i(\func_{\nu,\ell}\,\partial_\tau \func_{\nu,\ell}-\func_{\nu,\ell}\, \partial_\tau \func_{\nu,\ell} )\cosh^2(\tau)|_{\tau=\tau_0} \delta(\nu-\nu'),
\ee
which is manifestly \emph{zero}. Hence, with this choice of normalization, we get the canonical commutation relations for the creation and annihilation operators \eqref{btbtdagger}.

\subsubsection{Orthogonality relations}
Next, we will find the orthogonality relations satisfied by $\func_{\nu,\ell}(\tau)$. Recalling \eqref{ptolegp}, we see that we can use the orthogonality relations satisfied by associated Legendre polynomials with imaginary order are given by \cite{Hutasoit:2009xy,Bielski}
\be
\int_{-1}^{1} \frac{P_{\ell}^{i q}(x) P_{\ell}^{-i q'}(x)}{1-x^2} \, dx = 
\frac{2 \sinh(\pi q)}{q} \delta(q-q'),
\ee
where $q \in \mathbb{R}, \ell \in \mathbb{Z}$.
Using \eqref{ptolegp}, we find
\be
\int_{-\infty}^{\infty} p_{\nu,\ell}(\tau) p_{-\nu',\ell}(\tau) \, \cosh^2\tau \, d\tau = 
2 \pi  \,\,\delta(\nu-\nu').
\ee
Hence, we find that the Euclidean modes satisfy the following orthogonality relations (with $\nu,\nu'>0$)
\be
\begin{split}
\int_{-\infty}^{\infty} \func_{\nu',\ell}(\tau) \func^*_{\nu,\ell}(\tau) \, \cosh^2\tau \, d\tau 
&= 2 \pi (1+|c_{\nu,\ell}|^2) \delta(\nu-\nu')\\
\int_{-\infty}^{\infty} \func_{\nu',\ell}(\tau) \func_{\nu,\ell}(\tau) \, \cosh^2\tau \, d\tau 
&= 4 \pi \, c_{\nu,\ell} \,\delta(\nu-\nu').
\end{split}
\ee
Finally, we can define the function
\be
	\orthfunc_{\nu,\ell} (\tau) = {1+|c_{\nu,\ell}|^2 \over 2 \pi ( 1 - |c_{\nu,\ell}|^2)^2 } \, \func^*_{\nu, \ell} (\tau) - {c^{*}_{\nu,\ell} \over \pi (1 - |c_{\nu,\ell}|^2)^2 } \, \func_{\nu, \ell} (\tau),
\ee
such that, for $\nu > 0$ and $\nu' > 0$, 
\be
	\begin{split}
	\int_{-\infty}^{\infty} \orthfunc_{\nu',\ell}(\tau) \func_{\nu,\ell}(\tau) \, \cosh^2\tau \, d\tau  &= \delta(\nu -\nu'),\\
	\int_{-\infty}^{\infty} \orthfunc_{\nu',\ell}(\tau) \func^*_{\nu,\ell}(\tau) \, \cosh^2\tau \, d\tau  &= 0.
	\end{split}
\ee

\subsubsection{Transformation of conformal to spherical modes \label{conftospheranalyticity}}
We now determine the transformation between the spherical and the conformal mode functions. Since any function can be expanded in spherical harmonics we know that
\be
\label{confinsph}
{1 \over (-X\cdot\xi - i \epsilon)^{1+i\nu}} = \sum\limits_{\vec{\ell\,}} h_{\nu, \vec{\ell\,} , \xi} (\tau) Y_{\vec{\ell\, }}(\sph),
\ee
where on the right hand side $\tau, \Omega$ are related to $X$ via \eqref{xtotauomega}.

From symmetry considerations (see the discussion of transformation of plane waves to spherical waves in \cite{Sakurai:1995}) we conclude that
\be
h_{\nu, \vec{\ell}, \xi}(\tau) = h_{\nu, \ell}(\tau) Y_{\vl}(\hat{\xi}), 
\ee
where, as in the main text, we have $\xi = (1, \hat{\xi})$.

To determine the function $h_{\nu, \ell}(\tau)$ it suffices to consider the special case where $\hat{\xi} = (1, 0, 0)$ so that $Y_{\vl}(\hat{\xi}) = \sqrt{2 l + 1 \over 4 \pi } \delta_{0 m}$.  For this choice we also have
\be
-X \cdot \xi = \sinh \tau + \cosh \tau \cos \theta. 
\ee

Therefore, the sum in \eqref{confinsph} collapses to the case $m = 0$ and for this case $Y_{\ell}(\Omega) = \sqrt{2 l + 1 \over 4 \pi} P_{\ell}(\cos \theta)$ which is just a Legendre polynomial. By the orthogonality of Legendre polynomials, we have
\be
\label{hintegral}
h_{\nu, \ell}(\tau) = 2 \pi \int_{0}^{\pi} d \cos \theta {P_{\ell}(\cos \theta) \over (\sinh \tau + \cosh \tau \cos \theta - i \epsilon)^{1 + i \nu}}. 
\ee

The Legendre polynomial can be written as
\be
\label{legendreexpansion}
P_{\ell}(\cos \theta) = {1 \over 2^{\ell}} \sum_{k=0}^{[\ell/2]} (-1)^k {\ell \choose  k} {2 \ell - 2 k \choose \ell} (\cos \theta)^{\ell - 2 k}.
\ee
So to evaluate the expression above, we consider the auxiliary integral (obtained by substituting $x = \cos \theta$ above)
\be
\begin{split}
	q_{n, \nu}(\tau) &= \int_{-1}^{1} {x^n d x \over (\sinh \tau + \cosh \tau x - i \epsilon)^{1 + i \nu}} \\
	&=\left. {x^{n+1} \over n+1} \frac{1}{(\sinh\tau)^{1+i\nu}} \, _2F_1(n+1,1+i \nu ;n+2;-x \coth (\tau )) \right|_{-1}^{1}.
\end{split}
\ee
In the answer above, the $i \epsilon$ prescription has been absorbed into $\sinh \tau$ and we assume that $\sinh \tau$ has a small negative imaginary part.

The expression above can be parsed by performing a hypergeometric transformation (\cite[Eq.~15.8.20]{NIST:DLMF})
\be
\begin{split}
	_2F_1(a,b,c,z) = &{\Gamma(c) \pi  \csc (\pi  (-a-b+c))}  \\ \times \Big(&\frac{z^{-a} 
		\, _2F_1\left(a,a-c+1;a+b-c+1;1-\frac{1}{z}\right)}{\Gamma (c-a) \Gamma
		(c-b) \Gamma(a + b - c + 1)} 
	\\  &-\frac{z^{a-c} (1-z)^{-a-b+c}  \,   _2F_1\left(1-a,c-a;-a-b+c+1;1-\frac{1}{z}\right)}{\Gamma (a) \Gamma (b) \Gamma(-a - b + c + 1)} \Big).
\end{split}
\ee
Applying the above transformation,
and evaluating the indefinite integral at $x = 1$ and $x = -1$ 
we find, after some algebra,
\be
\label{finaltransform}
\begin{split}
	q_{n, \nu}(\tau) = { i \over \nu } \text{sech} (\tau )  \Big( 
	&(e^{\tau} - i \epsilon)^{-i \nu} \, _2F_1(1,-n;1-i \nu;1+\tanh \tau ) \\
	&+(-1)^{n+1}(-e^{-\tau} - i \epsilon)^{-i \nu}  \, _2F_1(1,-n;1-i \nu;1-\tanh\tau) \Big).
\end{split}
\ee

The hypergeometric functions that appear above are {\em polynomials} since we note that their second arguments are negative integers.  The $i \epsilon$ prescription is important since it tells us how to choose the branch of the power-function for real $\tau$. For real $\tau$ we have 
\be
\begin{split}
	q_{n, \nu}(\tau) = { i \over \nu } \text{sech} (\tau )  \Big( 
	&e^{i \nu \tau}  \, _2F_1(1,-n;1-i \nu;1+\tanh \tau ) \\
	&+(-1)^{n+1}e^{-\pi \nu} e^{- i \nu \tau}   \, _2F_1(1,-n;1-i \nu;1-\tanh\tau) \Big), \qquad \tau \in {\cal R}.
\end{split}
\ee

Substituting this expression in the expansion \eqref{legendreexpansion} we find, remarkably, that
\be
\label{hnuellsum}
h_{\nu, \ell} = 2 \pi {1 \over 2^{\ell}} \sum_{k=0}^{[\ell/2]} (-1)^k {\ell \choose  k} {2 \ell - 2 k \choose \ell} q_{\ell-2 k, \nu} = (2 \pi \nu) e^{i {\mathfrak b}(\nu, \ell)} \func_{\nu, \ell}(\tau), 
\ee
where the phase factor is given by
\be
\label{appphasefac}
e^{i {\mathfrak{b}(\nu, \ell)}} =  {1 \over i} \frac{(-1)^l (1-i \nu )_l}{(i \nu +1)_l}. 
\ee
Note that the factor of $2 \pi \nu$ in \eqref{hnuellsum} is precisely ${\tN_{\nu} \over N_{\nu}}$.

\section{Difficulty of representing massive particles at null infinity \label{appnulldiff}}

In this section, we review the difficulties in defining an extrapolate dictionary for massive fields at $\scrip$.   It is well known that future and past null infinity $\scri^{\pm}$ only admit massless representations of the Poincare group and this itself suggests that null infinity cannot be  used to obtain an algebra of boundary observables for massive fields \cite{Helfer:1993hv}.  Before we describe massive fields, we first review how null infinity provides a convenient home for massless field operators.

\subsection{Massless fields \label{edmfao}}
For massless scalar field, an extrapolate dictionary at ${\cal I}^{\pm}$ is obtained in a straightforward manner from the asymptotic expansion of the scalar field in the neighbourhood of ${\cal I}^{+}$ or ${\cal I}^{-}$ respectively.  Below, we describe the dictionary at $\scrip$ but an entirely analogous construction is valid at $\scrim$.

Consider a general asymptotically flat spacetime where as $r \rightarrow \infty$ the metric takes the form
\be
ds^2 = -du^2 + 2 du d r + r^2 \gamma_{AB} d x^{A} d x^{B}, 
\ee
where $\gamma_{AB}$ is the round metric on the celestial sphere and $u$ is the retarded time. 

With suitable boundary conditions at $u\, =\, -\, \infty$, a solution to the massless scalar field equation, with arbitrary interactions takes the asymptotic form
\be
\label{asymptexp}
\psi(u, r, x)\, =\, {1 \over r} \op(u, x) + \ldots
\ee
 The $\ldots$ in \eqref{asymptexp} contain subleading terms in $r$ but also terms  which may fall-off as $\frac{\ln r^{n}}{r^{n+1}}\, \vert\, n\, \geq\, 1$.\cite{Henneaux:2018mgn,Banerjee:2017jeg,Campiglia:2017dpg}\footnote{Such terms may be present if one has long range gravitational interaction or electromagnetic interaction in the case of charged scalar field.} In \eqref{asymptexp}, $\op(u, \hat{x})$ is free data and, together with the equations of motion, this data is sufficient to reconstruct the subleading terms in \eqref{asymptexp} and therefore obtain a solution away from the $r \rightarrow \infty$ limit.
For example, in the free field case in Minkowski space-time, one can set all the $\frac{\ln r}{r^n}$ terms to zero by choosing appropriate boundary conditions at $u = -\infty$ and determine the bulk field in terms of its boundary values as
\be
\label{flatsmearing}
\psi(r, u, x)\, =\, \int_{{\cal I}^{+}}\, d\tilde{u}\, d^{2} y \, {\cal K}_{m=0}(r, u, x\vert \tilde{u}, y)\, \op(\tilde{u}, y), 
\ee
where
\be
{\cal K}_{m=0}(r, u, \hat{x}\vert \tilde{u}, \hat{y})\, =\,  \frac{1}{[\, (\tilde{u} - u)\, +\, r(1 - \hat{x}\cdot \hat{y})\, ]}. 
\ee
Note that the smearing function ${\cal K}$ itself 
decays as $\frac{1}{\tilde{u}}$ as $\tilde{u}\, \rightarrow\, \pm\infty$.

The bulk massless field operator then defines an operator that is intrinsic to ${\scrip}$ via
\be
\label{extrapexample}
\lim_{r\, \rightarrow\, \infty}\, r\, \psi(u, r, \hat{x})\, \eqdef \, \op(u, \hat{x}). 
\ee

A salient property of this dictionary at $\scrip$ is that $\op(u, \hat{x})$ is a {\em free field operator} at ${\cal I}^{+}$. A $\star$-algebra can be defined using $\op(u, \hat{x})$ and the commutation relation
\be
\label{commutmasslesscal}
[\, \partial_{u}\op(u, \hat{x}),\, \partial_{u^{\prime}}\op(u^{\prime}, \hat{y})\, ]\, =\, \partial_{u}\delta(u - u^{\prime})\, \delta^{2}(\hat{x}, \hat{y}). 
\ee
This commutation relation is consistent with the extrapolate limit : The ``large-$r$" limit of a bulk commutator equals the commutator generated by the free field operator, $\op(u,\hat{x})$.   
\be
\lim_{r\, \rightarrow\, \infty}\, \lim_{r^{\prime}\, \rightarrow\, \infty}\, [\, \partial_{u}\psi(u, r, \hat{x}), \partial_{u^{\prime}}\psi(u^{\prime}, r^{\prime}, \hat{y})\, ]\, =\, \partial_{u}\delta(u - u^{\prime})\, \delta^{2}(\hat{x}, \hat{y}). 
\ee
We note that the extrapolate dictionary is a natural consequence of the fact that the total energy momentum flux of a massless field is non-zero and  finite at ${\cal I}^{\pm}$. The radiative flux that leaks at ${\cal I}^{+}$ can be computed using $\op(u, \hat{x})$.  

The commutation relations allow for the definition of a natural Fock space  at $\scrip$. We refer the reader to \cite{Laddha:2020kvp} for more details. Starting with the vacuum $|0 \rangle$, a Fock space is obtained through
\be
\label{h0sector}
{\cal H}^{0} = \text{span}\{\op(f_1) \ldots \op(f_n) | 0 \rangle  \}, 
\ee
where $\op(f_i)$ are obtained by smearing $\op(u, x)$ with smearing functions $f_i(u, x)$.

\subsection{Massive fields} 

We now explain why the steps above cannot easily be replicated for massive fields. Consider a simple free massive scalar field in Minkowski space. Using the standard partial wave expansion we find that near $r \rightarrow \infty$ the field can be written as
\be
\label{partialwave}
\phi(t, u, \Omega) \underset{r \rightarrow \infty}{\longrightarrow} \sum_{\ell,m} \int d k   {e^{i k r}  \over r} Y_{\vl}(\Omega) e^{-i \omega (u - r)} \phi_{k, \vl} + \hc,
\ee
where the mass-shell condition fixes $\omega^2 = k^2 + m^2$. 

Unlike the massless field since $\omega > r$, the expression above contains an oscillating piece $e^{-i(\omega - k) r}$. This means that each mode oscillates infinitely fast near $r \rightarrow \infty$ and, moreover, modes of different energies oscillate at different rates. Therefore there is no straightforward way to take a $r \rightarrow \infty$ limit for the field and obtain an algebra of observables at $\scrip$ or $\scrim$.

The expansion \eqref{partialwave} also makes it clear that if one considers classical solutions where $\phi_{k, \vl}$ varies smoothly with $k$, then the field near $r \rightarrow \infty$ dies off faster than any power of ${1 \over r}$ \cite{winicour1988massive}.

One might naively believe that for ``high-energy processes'', it is possible to neglect the mass of the field. But a good limit at $\scrip$ and $\scrim$ is only obtained for momenta that scale with $r$. This corresponds to decomposing the field as 
\be
\label{uvirdecomp}
\phi(u,r, \hat{x})\, =\, \phi_{<}(u, r, \hat{x}) + \phi_{>}(u, r, \hat{x}), 
\ee
where $\phi_{<}$ is obtained by integrating over Fourier modes  with $\vert k\vert\, \leq\, m^{2}r$, and $\phi_{>}$ is composed of the ultra-violet modes with $\vert k \vert\, \geq\, m^{2}r$.  The split of $\phi$ between $\phi_{>},\, \phi_{<}$ is not covariant and relies on a choice of frame. In such a frame, it can be checked that $\phi_{>}$ has a power law fall-off as $r\, \rightarrow\, \infty$
\be
\phi_{>}(u, r, \hat{\Omega})\, =\,  \frac{1}{r}\, \int_{\pi m^{2} r}^{\infty}\, [\, \phi_{\vert k\vert, \hat{\Omega}} e^{-i \vert k\vert u}\, + \hc] d k + O(\frac{1}{r^{2}}). 
\ee
On the other hand, $\phi_{<}$ decays faster than any power of $r$.

The decomposition does not yield a natural extrapolated algebra of observables at null infinity since the definition of $\phi_{>}$ is, itself, dependent on $r$. Such a  non-covariant decomposition of the massive field  in ``heavy" and ``light" fields may still be useful in some situations. One example of such a process is a $2\, \rightarrow\, 2$ gravitational scattering involving minimally coupled scalars with an impact parameter $\vec{b}$. In the COM frame with relative momentum $2\, \vec{p}$ , this process can be analysed in a space-time region bounded by a time-like hyperboloid at a distance $\rho\, \sim\, \frac{\vert p\vert}{m^{2}}$. In the eikonal limit, where $\rho\, \rightarrow\, \infty$ keeping the momentum transfer fixed, the $\phi_{>}$ modes  dominantly contribute to the outgoing states.

We refer the reader to \cite{Dappiaggi:2007hh} for an interesting attempt where a massive field is represented by a pair of auxiliary massless fields at ${\cal I}$. We believe that the issue of whether an innovative construction might yield an extrapolated set of observables for massive fields on ${\cal I}$ deserves attention.

\section{Sample two-point calculations}
In this appendix, we present some sample two-point computations.  We work in the s-wave sector where the spherical modes simplify.

\subsection{Boundary two-point function}
First, we will compute the two point function of two boundary operators smeared uniformly over the sphere on $\spatinf$ and smeared with a Gaussian function in de Sitter time.  More specifically, we study the operator $\opspat(\gfull)$ where
\be
\gfull(\tau, \Omega) = \sqrt{\pi \over a} e^{-a \tau^2} \text{sech}(\tau), \quad \text{independent~of~$\Omega$},
\ee
and compute $\langle 0 | \opspat(\gfull)^2 | 0 \rangle$.

We first note that
\be
\int \opspat(\tau, \Omega) d \Omega =
\sqrt{4 \pi} \int_0^{\infty} d \nu \tN_{\nu}  \tB_{\nu, 0} \func_{\nu,0}(\tau).
\ee
The factor of $\sqrt{4 \pi}$ arises from combining the area of the sphere ($4 \pi$) with the uniform value of the zeroth spherical harmonic ${1 \over \sqrt{4 \pi}}$ normalized according to \eqref{normsphericalharm}. 
We also note that
\be
\sqrt{4 \pi} \tN_{\nu} \func_{\nu,0}(\tau)  = \sqrt{2 \over  \pi} {1 \over \cosh \tau} \left[e^{\pi \nu \over 2} e^{-i \nu \tau} - e^{-{\pi \nu \over 2}} e^{i \nu \tau} \right],
\ee
and so we find that
\be
\tN_{\nu} \gfull_{\nu, 0} = \tN_{\nu} \int d \Omega d \tau \cosh^2 \tau \gfull(\tau, \Omega) \func_{\nu, 0}(\tau) = {\sqrt{8 \pi} \over a} e^{-\nu^2 \over 4 a} \sinh({\pi \nu \over 2})
\ee
in the notation of \eqref{gfullexpansion}.

Therefore the two point function of the smeared operator in the vacuum is
\be
\langle 0| \opspat(g) \opspat(\gfull) | 0 \rangle = \int \tN_{\nu}^2 |g_{\nu, 0}|^2 d \nu =\left({2 \pi \over a} \right)^{3 \over 2}  \left(e^{\frac{\pi ^2 a}{2}}-1\right). 
\ee

\subsection{Bulk to boundary two-point function}
Next, we rederive the bulk to boundary two-point function using the mode expansion for the bulk field and the boundary operator.

First, let us derive the expected result. We know that the two point function of bulk fields in the vacuum is
\be
\langle 0 | \phi(\rho_1, X_1) \phi(\rho_2, X_2) |0 \rangle = {m \over 4 \pi^2} {K_1(m w) \over w}, 
\ee
with 
\be
w = \sqrt{\rho_1^2 + \rho_2^2 - 2 \rho_1 \rho_2 X_1 \cdot X_2}. 
\ee

From here we find that
\be
\langle \opspat( X_1) \phi(\rho_2, X_2) \rangle = {m \over 4 \pi^2} e^{m \rho_2 X_1 \cdot X_2}. 
\ee
We now integrate both operators over the celestial sphere so as to get the s-wave two point function. We note that
\be
X_1 \cdot X_2 = -\sinh \tau_1 \sinh \tau_2 + \cosh \tau_1 \cosh \tau_2 \cos \theta_{1 2},
\ee
and therefore
\be
\begin{split}
&{m \over 4 \pi^2} \int e^{m \rho_2 X_1 \cdot X_2} d \Omega_1 d \Omega_2 = -2 m  e^{-m \rho_2 \sinh \tau_1 \sinh \tau_2}  \int_{-1}^{1}  e^{m \rho_2 \cosh \tau_1 \cosh \tau_2 \cos \theta} d (\cos \theta) \\ &= {-2  \over \rho_2 \cosh \tau_1 \cosh \tau_2} e^{-m \rho_2 \sinh \tau_1 \sinh \tau_2} \left(e^{m \rho_2 \cosh \tau_1 \cosh \tau_2} - e^{-m \rho_2 \cosh \tau_1 \cosh \tau_2} \right)\\ &= {2 \over \rho_2 \cosh \tau_1 \cosh \tau_2} \left(e^{-m \rho_2 \cosh(\tau_1 + \tau_2)} - e^{m \rho_2 \cosh(\tau_1 - \tau_2)}  \right).
\end{split}
\ee

Let us try and reproduce this using our mode expansion. In the spherical basis, we have
\be
\begin{split}
&\int \opspat(X_1) d \Omega_1 = \sqrt{4 \pi} \int d \nu \tB_{\nu, 0} \tN_{\nu} \func_{\nu, 0}(\tau_1) + \hc\\
&= \sqrt{2 \over \pi} \int d \nu \tB_{\nu, 0} {1 \over  \cosh(\tau_1)} \left[e^{\pi \nu \over 2} e^{-i \nu \tau_1} - e^{-{\pi \nu \over 2}} e^{i \nu \tau_1} \right] + \hc
\end{split}
\ee
We also find that
\be
\begin{split}
\int \phi(\rho_2, X_2) d \Omega_2 &= \sqrt{4 \pi}  \int \tB^{\dagger}_{\nu, 0} {K_{i \nu}(m \rho_2) \over \rho_2} \tN_{\nu} \func_{\nu, 0}(\tau_2) + \hc\\
&= \sqrt{2 \over \pi} \int d \nu \tB_{\nu, 0}^{\dagger} {K_{i \nu}(m \rho_2)  \over \rho_2} {1 \over  \cosh(\tau_2)} \left[e^{\pi \nu \over 2} e^{i \nu \tau_2} - e^{-{\pi \nu \over 2}} e^{-i \nu \tau_2} \right].
\end{split}
\ee
This tells us that
\be
\label{zphiintegans}
\begin{split}
\int \langle \opspat(X_1) \phi(\rho, X_2) \rangle d \Omega_1 d \Omega_2 = &{4 \over \pi \rho_2 \cosh \tau_1 \cosh \tau_2}  \\ &\times \int_{0}^{\infty} d \nu  K_{i \nu}(m \rho_2) \left[\cos({i \nu (\tau_1 + \tau_2 - {i \pi})}) - \cos({i \nu (\tau_1 - \tau_2)})\right].
\end{split}
\ee
This integral is divergent at large $\nu$. This is the same divergence that we have already encountered multiple times. Therefore \eqref{zphiintegans} should be understood as follows: we first integrate both sides with an appropriate function of $\tau_1, \tau_2$ --- such as a Gaussian in $\tau_1$ and $\tau_2$ --- that damps the large $\nu$-divergence and subsequently perform the $\nu$-integral. 

This process can be short-circuited using the following formal manipulation. We first note the Fourier transform for the Macdonald function which we have already used above (chapter 12 of \cite{Erdelyi:1954:TI2})
\be
\label{macdonaldft}
\int_{0}^{\infty} K_{i \nu}(x) \cos(\nu t) {d \nu \over 2 \pi} =  {1 \over 4} e^{-x \cosh t}.
\ee
To apply this formula to \eqref{zphiintegans}, we simply set $t = \tau_1 + \tau_2 - i \pi$ and note that $\cosh(\tau_1 + \tau_2 - i \pi) = -\cosh(\tau_1 + \tau_2)$. Then we write,
\be
\int_0^{\infty} K_{i \nu}(m \rho_2) \cos({i \nu (\tau_1 + \tau_2 - {i \pi})}) d \nu \rightarrow {\pi \over 2} e^{m \rho_2 \cosh(\tau_1 + \tau_2)},
\ee
with the understanding that equality is only obtained when both sides are integrated with an appropriate function of $\tau_1, \tau_2$.

With this step, we find
\be
\int \langle \opspat(X_1) \phi(\rho, X_2) \rangle d \Omega_1 d \Omega_2 = {2 \over \rho_2 \cosh \tau_1 \cosh \tau_2}   \left[e^{-m \rho_2 \cosh{(\tau_1 + \tau_2)}} - e^{m \rho_2 \cosh{(\tau_1 - \tau_2)}} \right], 
\ee
in perfect agreement with the direct expression obtained above.

\bibliographystyle{utphys}
\bibliography{references}
\end{document}